\documentclass{pasj00}

\def\commenta{$^*$}
\def\commentb{$^\dagger$}
\def\commentc{$^\ddagger$}
\def\commentd{$^\S$}
\def\commente{$^\|$}
\def\commentf{$^\#$}

\def\inpress{in press}

\def\arxiv#1{ (arXiv astro-ph/#1)}

\DeclareAbbreviation\AAHam{Astron. Abh. Hamburg. Sternw.}
\DeclareAbbreviation\AARv{Astron. Astrophys. Rev.}
\DeclareAbbreviation\AAS{American Astron. Soc. Meeting Abstracts}
\DeclareAbbreviation\AcA{Acta Astron.}
\DeclareAbbreviation\actaa{Acta Astron.}
\DeclareAbbreviation\Afz{Astrofizika}
\DeclareAbbreviation\AGAb{Astronomische Gesellschaft Abstract Ser.}
\DeclareAbbreviation\an{Astron. Nachr.}
\DeclareAbbreviation\AnAp{Annales d'Astrophysique}
\DeclareAbbreviation\AnTok{Tokyo Astron. Obs. Annals, Sec. Ser.}
\DeclareAbbreviation\Ap{Astrophysics}
\DeclareAbbreviation\ARep{Astron. Rep.}
\DeclareAbbreviation\AstBu{Astrophys. Bull.}
\DeclareAbbreviation\ATel{Astron. Telegram}
\DeclareAbbreviation\ATsir{Astron. Tsirk.}
\DeclareAbbreviation\AcApS{Acta Astrophys. Sinica}
\DeclareAbbreviation\AstL{Astron. Lett.}
\DeclareAbbreviation\BaltA{Baltic Astron.}
\DeclareAbbreviation\BANS{Bull. of the Astron. Institutes of the Netherlands Suppl. Ser.}
\DeclareAbbreviation\BASI{Bull. Astron. Soc. India}
\DeclareAbbreviation\BeSN{Be Newslett.}
\DeclareAbbreviation\BHarO{Harvard Coll. Obs. Bull.}
\DeclareAbbreviation\CBET{Cent. Bur. Electron. Telegrams}
\DeclareAbbreviation\ChJAA{Chinese J. of Astron. and Astrophys.}
\DeclareAbbreviation\caa{Chinese J. of Astron. and Astrophys.}
\DeclareAbbreviation\CoAsi{Asiago Contr.}
\DeclareAbbreviation\CoSka{Contributions of the Astronomical Observatory Skalnat\'e Pleso}
\DeclareAbbreviation\GCN{GRB Coord. Netw. Circ.}
\DeclareAbbreviation\ErgAN{Erg. Astron. Nachr.}
\DeclareAbbreviation\ibvs{IBVS}
\DeclareAbbreviation\IEEEP{IEEE Proc.}
\DeclareAbbreviation\JAD{J. Astron. Data}
\DeclareAbbreviation\JAVSO{J. American Assoc. Variable Star Obs.}
\DeclareAbbreviation\JBAA{J. Br. Astron. Assoc.}
\DeclareAbbreviation\JPhCS{J. of Physics Conference Series}
\DeclareAbbreviation\JPSJ{J. Phys. Soc. Japan}
\DeclareAbbreviation\JSARA{J. of the Southeastern Assoc. for Research in Astron.}
\DeclareAbbreviation\LowOB{Lowell Obs. Bull.}
\DeclareAbbreviation\MitAG{Mitteil. der Astronom. Gesell. Hamburg}
\DeclareAbbreviation\MitVS{Mitteil. Ver\"{a}nderl. Sterne}
\DeclareAbbreviation\MmSAI{Mem. Soc. Astron. Ital.}
\DeclareAbbreviation\memsai{Mem. Soc. Astron. Ital.}
\DeclareAbbreviation\Msngr{Messenger}
\DeclareAbbreviation\NewA{New Astron.}
\DeclareAbbreviation\na{New Astron.}
\DeclareAbbreviation\NewAR{New Astron. Rev.}
\DeclareAbbreviation\nar{New Astron. Rev.}
\DeclareAbbreviation\NInfo{Nauchnye Informatsii}
\DeclareAbbreviation\OAP{Odessa Astron. Publ.}
\DeclareAbbreviation\Obs{Observatory}
\DeclareAbbreviation\OEJV{Open Eur. J. on Variable Stars}
\DeclareAbbreviation\PASA{Publ. Astron. Soc. Australia}
\DeclareAbbreviation\PASAu{Publ. Astron. Soc. Australia}
\DeclareAbbreviation\PAZh{Pis'ma AZh}
\DeclareAbbreviation\POBeo{Publ. de l'Observatoire Astronomique de Beograd}
\DeclareAbbreviation\PCCP{Phys. Chem. Chem. Phys.}
\DeclareAbbreviation\PhR{Phys. Rep.}
\DeclareAbbreviation\PVSS{Publ. Variable Stars Sect. R. Astron. Soc. New Zealand}
\DeclareAbbreviation\PZ{Perem. Zvezdy}
\DeclareAbbreviation\PZP{Perem. Zvezdy, Prilozh.}
\DeclareAbbreviation\QJRAS{QJRAS}
\DeclareAbbreviation\RA{Ricerche Astronomiche}
\DeclareAbbreviation\RMxAA{Rev. Mexicana Astron. Astrof.}
\DeclareAbbreviation\RvMA{Reviews of Modern Astron.}
\DeclareAbbreviation\SASS{Society for Astronom. Sciences Ann. Symp.}
\DeclareAbbreviation\Sci{Science}
\DeclareAbbreviation\SPIE{SPIE Proc.}
\DeclareAbbreviation\SvA{Soviet Astronomy}
\DeclareAbbreviation\SvAL{Soviet Astronomy Letters}
\DeclareAbbreviation\VeSon{Ver\"{o}ff. Sternw. Sonneberg}
\DeclareAbbreviation\VSOLJBul{VSOLJ Variable Star Bull.}
\DeclareAbbreviation\yCat{VizieR Online Data Catalog}
\DeclareAbbreviation\ZA{Z. Astrophys.}

\def\ASPConf#1#2{ASP Conf. Ser. #1, #2}

\def\PublisherASP{San Francisco: ASP}

\newcounter{author}
\def\authorcount#1#2{\refstepcounter{author}\label{#1}
                     \altaffiltext{\ref{#1}}{#2}}

\begin{document}
\SetRunningHead{C. Nakata et al.}{MASTER OT J211258.65$+$242145.4 and MASTER OT J203749.39+552210.3}

\Received{201X/XX/XX}
\Accepted{201X/XX/XX}

\title{WZ Sge-type dwarf novae with multiple rebrightenings: MASTER OT J211258.65$+$242145.4 and MASTER OT J203749.39+552210.3}

\author{Chikako~\textsc{Nakata},\altaffilmark{\ref{affil:Kyoto}*}
        Tomohito~\textsc{Ohshima},\altaffilmark{\ref{affil:Kyoto}}
        Taichi~\textsc{Kato},\altaffilmark{\ref{affil:Kyoto}}
        Daisaku~\textsc{Nogami},\altaffilmark{\ref{affil:HidaKwasan}}
        Gianluca~\textsc{Masi},\altaffilmark{\ref{affil:Masi}} 
        Enrique~de~\textsc{Miguel},\altaffilmark{\ref{affil:Miguel}}$^,$\altaffilmark{\ref{affil:Miguel2}}
        Joseph~\textsc{Ulowetz},\altaffilmark{\ref{affil:Ulowetz}}
        Colin~\textsc{Littlefield},\altaffilmark{\ref{affil:LCO}}
        William~N.~\textsc{Goff},\altaffilmark{\ref{affil:Goff}}
        Thomas~\textsc{Krajci},\altaffilmark{\ref{affil:Krajci}}
        Hiroyuki~\textsc{Maehara},\altaffilmark{\ref{affil:Kiso}}
         William~\textsc{Stein},\altaffilmark{\ref{affil:Stein}}
         Richard~\textsc{Sabo},\altaffilmark{\ref{affil:Sabo}}
         Ryo~\textsc{Noguchi},\altaffilmark{\ref{affil:OKU}}
         Rikako~\textsc{Ono},\altaffilmark{\ref{affil:OKU}}
         Miho~\textsc{Kawabata},\altaffilmark{\ref{affil:OKU}}
         Hisami~\textsc{Furukawa},\altaffilmark{\ref{affil:OKU}}
         Katsura~\textsc{Matsumoto},\altaffilmark{\ref{affil:OKU}}
         Takehiro~\textsc{Ishibashi},\altaffilmark{\ref{affil:OKU}}
         Pavol~A.~\textsc{Dubovsky},\altaffilmark{\ref{affil:Dubovsky}}
         Igor~\textsc{Kudzej},\altaffilmark{\ref{affil:Dubovsky}}
         Shawn~\textsc{Dvorak},\altaffilmark{\ref{affil:Dvorak}}
        Franz-Josef~\textsc{Hambsch},\altaffilmark{\ref{affil:GEOS}}$^,$\altaffilmark{\ref{affil:BAV}}$^,$\altaffilmark{\ref{affil:Hambsch}} 
        Roger~D.~\textsc{Pickard},\altaffilmark{\ref{affil:BAAVSS}}$^,$\altaffilmark{\ref{affil:Pickard}}
        Etienne~\textsc{Morelle},\altaffilmark{\ref{affil:Morelle}}
        Eddy~\textsc{Muyllaert},\altaffilmark{\ref{affil:VVSBelgium}}
        Stefano~\textsc{Padovan},\altaffilmark{\ref{affil:AAVSO}}
        Arne~\textsc{Henden},\altaffilmark{\ref{affil:AAVSO}}
}

\authorcount{affil:Kyoto}{
     Department of Astronomy, Kyoto University, Kyoto 606-8502}
\email{$^*$nakata@kusastro.kyoto-u.ac.jp}

\authorcount{affil:HidaKwasan}{
     Kwasan Observatory, Kyoto University, Yamashina,
     Kyoto 607-8471}

\authorcount{affil:Masi}{
     The Virtual Telescope Project, Via Madonna del Loco 47, 03023
     Ceccano (FR), Italy}

\authorcount{affil:Miguel}{
     Departamento de F\'isica Aplicada, Facultad de Ciencias
     Experimentales, Universidad de Huelva,
     21071 Huelva, Spain}

\authorcount{affil:Miguel2}{
     Center for Backyard Astrophysics, Observatorio del CIECEM,
     Parque Dunar, Matalasca\~nas, 21760 Almonte, Huelva, Spain}

\authorcount{affil:Ulowetz}{
     Center for Backyard Astrophysics Illinois,
     Northbrook Meadow Observatory, 855 Fair Ln, Northbrook,
     Illinois 60062, USA}

\authorcount{affil:LCO}{
     Department of Physics, University of Notre Dame, Notre Dame,
     Indiana 46556, USA}

\authorcount{affil:Goff}{
     13508 Monitor Ln., Sutter Creek, California 95685, USA}

\authorcount{affil:Krajci}{
     Astrokolkhoz Observatory,
     Center for Backyard Astrophysics New Mexico, PO Box 1351 Cloudcroft,
     New Mexico 83117, USA}

\authorcount{affil:Kiso}{Kiso Observatory, Institute of Astronomy,
 School of Science, The University of Tokyo,
10762-30, Mitake, Kiso-machi, Kiso-gun, Nagano 397-0101, Japan}

\authorcount{affil:Stein}{
     6025 Calle Paraiso, Las Cruces, New Mexico 88012, USA}

\authorcount{affil:Sabo}{
     2336 Trailcrest Dr., Bozeman, Montana 59718, USA}

\authorcount{affil:OKU}{
     Osaka Kyoiku University, 4-698-1 Asahigaoka, Osaka 582-8582}

\authorcount{affil:Dubovsky}{
     Vihorlat Observatory, Mierova 4, Humenne, Slovakia}

\authorcount{affil:Dvorak}{
     Rolling Hills Observatory, 1643 Nightfall Drive,
     Clermont, Florida 34711, USA}

\authorcount{affil:GEOS}{
     Groupe Europ\'een d'Observations Stellaires (GEOS),
     23 Parc de Levesville, 28300 Bailleau l'Ev\^eque, France}

\authorcount{affil:BAV}{
     Bundesdeutsche Arbeitsgemeinschaft f\"ur Ver\"anderliche Sterne
     (BAV), Munsterdamm 90, 12169 Berlin, Germany}

\authorcount{affil:Hambsch}{
     Vereniging Voor Sterrenkunde (VVS), Oude Bleken 12, 2400 Mol, Belgium}

\authorcount{affil:BAAVSS}{
     The British Astronomical Association, Variable Star Section (BAA VSS),
     Burlington House, Piccadilly, London, W1J 0DU, UK}

\authorcount{affil:Pickard}{
     3 The Birches, Shobdon, Leominster, Herefordshire, HR6 9NG, UK}

\authorcount{affil:Morelle}{
     9 rue Vasco de GAMA, 59553 Lauwin Planque, France}

\authorcount{affil:VVSBelgium}{
     Vereniging Voor Sterrenkunde (VVS),  Moffelstraat 13 3370
     Boutersem, Belgium}

\authorcount{affil:AAVSO}{
     American Association of Variable Star Observers, 49 Bay State Rd.,
     Cambridge, MA 02138, USA}



\maketitle

\begin{abstract}
We report on  photometric observations of WZ Sge-type dwarf novae,
 MASTER OT J211258.65$+$242145.4 and MASTER OT J203749.39+552210.3
 which underwent outbursts in 2012. 
 Early superhumps were recorded in both systems.
 During superoutburst plateau, ordinary superhumps
 with a period of 0.060291(4) d
 (MASTER J211258) and of 0.061307(9) d (MASTER J203749) 
 in average were observed.
 MASTER J211258 and MASTER J203749 exhibited eight and more than
 four post-superoutburst rebrightenings, respectively.
 In the final part of the superoutburst, an increase in the
 superhump periods was seen in both systems.
 We have made a survey of WZ Sge-type
 dwarf novae with multiple rebrightenings, and confirmed that
 the superhump periods of WZ Sge-type dwarf novae
 with multiple rebrightenings were longer than those
 of WZ Sge-type dwarf novae without a rebrightening.
 Although WZ Sge-type dwarf novae with multiple rebrightenings have been
 thought to be the good candidates for
 period bouncers based on their low mass ratio
 ($q$) from inferred from the period of fully grown (stage B) superhumps,
 our new method using the period of growing superhumps
 (stage A superhumps), however, implies higher $q$ than those expected from
 stage B superhumps.  These $q$ values 
 appear to be consistent with
 the duration of the stage A superoutbursts, which likely reflects
 the growth time
 of the 3:1 resonance.
 We present a working hypothesis that the small fractional 
 superhump excesses for stage B superhumps in these systems
 may be explained as a result that a higher gas pressure effect
 works in these systems than in ordinary SU UMa-type dwarf novae.
 This result leads to a new picture that WZ Sge-type dwarf novae with
 multiple rebrightenings and SU UMa-type dwarf novae without
 a rebrightening (they are not period bouncers)
 are located in the same place on the evolutionary track.
\end{abstract}

\section{Introduction}

   Cataclysmic variables (CVs) are binary star systems that have a white dwarf (primary)
 and a secondary which fills its Roche lobe and transfers matter to the primary.  

   Dwarf novae (DNe) are one of subtypes of CVs. DNe have outbursts that
 are well understood as a release of gravitational energy caused by large mass
 transfer through the disk by the thermal instability.  SU UMa-type dwarf novae
 are a subclass of DNe.
They have occasional superoutbursts that are brighter and have longer durations
 than normal outbursts. During superoutbursts, they show light variations, which have 
a period few percent longer than the orbital period, called superhumps.
It is believed that superhumps are caused by the tidal instability that is triggered
 when the disk radius expands to the critical radius for the 3:1 resonance
 [see e.g. \citet{osa96review} for a theoretical review]. 
According to \citet{Pdot}, 
SU UMa-type dwarf novae generally show three distinct stages of period
 variation of superhumps; the first stage with a longer superhump period (stage A),
 middle stage with systematically varying periods (stage B)
 and final stage with a shorter superhump period (stage C).

WZ Sge-type dwarf novae are a subgroup of dwarf novae [see e.g. \cite{bai79wzsge};
 \cite{dow90wxcet}; \cite{kat01hvvir}]. They are known as systems that have
 infrequent large-amplitude superoutbursts. Although the general properties
 of outbursts in
 WZ Sge-type dwarf novae can be understood with thermal-tidal
 disk-instability model [see e.g. \cite{osa95wzsge}, \citet{osa03DNoutburst}
 for WZ Sge],
 there remain features in WZ Sge-type dwarf novae whose origin 
is still in dispute.

 WZ Sge-type dwarf novae have several characteristic properties.
 One is the existence of double-wave early superhumps with periods
 close to the orbital periods in early stage of superoutbursts
\citep{kat02wzsgeESH}. \citet{pat81wzsge} originally suggested 
that these humps represent enhanced orbital humps arising from 
an enhanced mass transfer.  \citet{osa02wzsgehump} suggested that
these humps arise from the 2:1 resonance. Although 
\citet{pat02wzsge} now appears to favor the explanation by
\citet{osa02wzsgehump} for the origin of early superhumps,
 \citet{pat02wzsge} suggested that
an enhanced mass transfer plays a role in WZ Sge-type outbursts --
post-superoutburst rebrightenings.
Post-superoutburst rebrightenings [also called ``echo outburst''
by \citet{pat02wzsge}] are often seen after superoutbursts
of WZ Sge-type dwarf novae.
These rebrightenings are classified by their morphology
 (\cite{ima06tss0222}; \cite{Pdot}).
The mechanism of rebrightenings is still in dispute. There is a suggestion that
 the enhanced mass-transfer following the superoutburst cause rebrightenings
 (\cite{ham00DNirradiationreview}; \cite{bua02suumamodel}).
 On the other hand, it is suggested that persistence of
 high viscosity in the accretion disk after the termination of
the superoutburst produce rebrightenings 
(\cite{osa97egcnc}, \cite{osa01egcnc}).
\citet{pat02wzsge} wrote ``hot-spot eclipses establish that 
mass transfer is greatly enhanced during superoutburst. 
This may settle the debate over the origin of echo outbursts 
in dwarf novae'', while \citet{osa03DNoutburst} reported
that the conclusion by \citet{pat02wzsge} was a result of
mis-interpretation of the observation, and there is no evidence
of an enhanced mass transfer.

 According to the standard evolutionary theory, CVs evolve from a longer to
 shorter orbital period and finally reach a minimum orbital period when the secondary
 begins to be degenerate or the thermal and mass-loss time scales
 become comparable for the secondary.
 Then the secondary becomes a brown dwarf which cannot
 remain in hydrogen burning or becomes somewhat oversized for its mass as
 a result of deviation from thermal equilibrium 
 and the orbital period becomes longer again.
 The systems
 whose periods increase after reaching the minimum period are called period
 bouncers [see e.g. \citet{kni11CVdonor} for standard evolutionary
 theory of CVs].
 A fraction of known WZ Sge-type dwarf novae are thought to be period
 bouncers because of their short orbital periods (\cite{uem10DNshortP};
 \cite{pat05j1255}; \cite{pat11CVdistance}).

Although the mechanism of rebrightenings are unknown, the rebrightenings
 have been thought to be related with evolutionary stages.
\citet{ima06tss0222} and \citet{Pdot} classified 
the rebrightenings of WZ Sge-type dwarf novae
 into four types by their light curve shapes: long duration
 rebrightenings (type-A), multiple rebrightenings (type-B), single
 rebrightening (type-C), and no rebrightening (type-D). 
  \citet{Pdot} indicated there is a relation between these
 rebrightening types and
$P_{\rm dot}$ versus $P_{\rm SH}$ [see figure 37 of \citet{Pdot}].
 This result implies that rebrightening type generally reflects 
the system parameters (the nature or the evolutionary state)
of the system, although different superoutbursts in the same
 system sometimes show different rebrightening types
[AL Com: \citet{uem08alcom}; WZ Sge: \citet{pat81wzsge}].
In EZ Lyn, however, two superoutbursts that have been observed so far
 were with type-B rebrightenings (\cite{pav07j0804}; 
\cite{kat09j0804}; \cite{Pdot3}), suggesing that the same
object tends to show the same type of rebrightenings.
The cases were also true for UZ Boo (\cite{kuu96TOAD}; \cite{Pdot})
and WZ Sge (1978 and 2011, \cite{pat81wzsge}; \cite{pat02wzsge};
\cite{ish02wzsgeletter}).  We here assume as the starting point
that WZ Sge-type dwarf novae can be generally categorized
by the rebrightening type.

EG Cnc, one of WZ Sge-type dwarf novae, is known to have had
a superoutburst with multiple rebrightenings (\cite{pat98egcnc};
 \cite{kat04egcnc}). It was
 discovered with its outbursts in 1977 by \citet{hur83egcnc}.
 {\citet{pat98egcnc}} calculated its mass ratio $q=0.027$ from the
 fractional superhump excess, defined as $\epsilon \equiv (P_{\rm sh}-P_{\rm orb})/P_{\rm orb}$.
 Using an average
 white dwarf mass of 0.7 $M_{\odot}$, this suggests very low secondary
 mass $M_2\approx 0.02M_{\odot}$.
After then, it has been generally considered that objects with
multiple rebrightenings are good candidates of period bouncers.

   In this paper, we present two WZ Sge-type dwarf novae which exhibited
 multiple rebrightenings. 
MASTER OT J211258.65$+$242145.4 was discovered by
the Mobile Astronomical System of the Telescope-Robots
(MASTER; \cite{MASTER}) network
(\cite{den12j2112atel4208}, hereafter MASTER J211258).
The quiescent counterpart was identified with a 20-th
magnitude star.  The large outburst amplitude ($\sim$7 mag) 
was suggestive of a WZ Sge-type dwarf nova (vsnet-alert 14697).
 MASTER OT J203749.39+552210.3 was discovered by MASTER
 network (\cite{bal12j2037atel4515},
hereafter MASTER J203749).  No previous outburst is known.
The quiescent counterpart was identified with a 20-th
magnitude star.
The large outburst amplitude (more than 7 mag) was again
suggestive of a WZ Sge-type dwarf nova.  These objects
turned out to be WZ Sge-type dwarf novae with multiple
rebrightenings.  We report in this paper on the development
of their outbursts and superhumps and discuss the implications
from the recorded variation of superhump periods.
In Section \ref{sec:obs}, we briefly show a log of
 observations and the analysis method. In Section
\ref{sec:res2112} and Section \ref{sec:res2037}, we present
 the results of the observations of MASTER J211258 and
 MASTER J203749, respectively.
 In Section \ref{sec:discussion}, we make a discussion
on the results.

\section{Observation and Analysis}\label{sec:obs}

\begin{table}
\caption{Log of observations of MASTER J211258.}\label{tab:log}
\begin{center}
\begin{tabular}{ccccccc}
\hline
Start\commenta & End\commenta & mag\commentb & error\commentc & $N$\commentd & obs\commente & sys\commentf \\
\hline
4.4570 & 4.5690 & 13.855 & 0.004 & 203 & deM & C \\
4.5147 & 4.6119 & 13.815 & 0.002 & 308 & Mas & C \\
4.6663 & 4.8883 & 13.584 & 0.002 & 235 & UJH & C \\
4.6923 & 4.8649 & 13.698 & 0.003 & 96 & BJA & V \\
4.6932 & 4.8818 & 13.907 & 0.001 & 664 & LCO & C \\
4.7362 & 4.8962 & 13.891 & 0.001 & 287 & SRI & C \\
4.7637 & 4.9541 & 13.902 & 0.001 & 458 & SWI & V \\
5.3734 & 5.5402 & 13.544 & 0.001 & 419 & DPV & C \\
5.4514 & 5.5950 & 13.998 & 0.001 & 483 & Mas & C \\
5.6468 & 5.8412 & 14.061 & 0.001 & 425 & LCO & C \\
\hline
  \multicolumn{7}{l}{\commenta BJD$-$2456100.} \\
  \multicolumn{7}{l}{\commentb Mean magnitude.} \\
  \multicolumn{7}{l}{\commentc 1-$\sigma$ of the mean magnitude.} \\
  \multicolumn{7}{l}{\commentd Number of observations.} \\
  \multicolumn{7}{l}{\commente Observer's code.} \\
  \multicolumn{7}{l}{\commentf Filter.  ``C'' means no filter (clear).} \\
\end{tabular}
\end{center}
\end{table}

\begin{table}
\caption{Log of observations of MASTER J203749.}\label{tab:log_j2037}
\begin{center}
\begin{tabular}{ccccccc}
\hline
Start\commenta & End\commenta & mag\commentb & error\commentc & $N$\commentd & obs\commente & sys\commentf \\
\hline
25.1165 & 25.2574 & 15.166 & 0.001 & 166 & Mas & C \\
25.9345 & 26.1221 & 15.171 & 0.002 & 445 & KU & C \\
25.9575 & 26.1298 & 15.236 & 0.003 & 441 & Mhh & C \\
26.1294 & 26.3079 & 15.302 & 0.002 & 207 & Mas & C \\
26.9299 & 27.1284 & 15.318 & 0.002 & 486 & KU & C \\
28.2880 & 28.4762 & 15.308 & 0.002 & 240 & deM & C \\
28.6062 & 28.7205 & 15.184 & 0.003 & 141 & Kra & C \\
29.2833 & 29.4669 & 15.338 & 0.004 & 230 & deM & C \\
29.5441 & 29.7184 & 15.247 & 0.004 & 167 & Kra & C \\
29.9108 & 30.0583 & 15.270 & 0.005 & 304 & KU & C \\
30.5411 & 30.7152 & 15.259 & 0.005 & 167 & Kra & C \\
31.5406 & 31.7128 & 15.326 & 0.004 & 165 & Kra & C \\
32.5403 & 32.7101 & 15.420 & 0.003 & 163 & Kra & C \\
33.5390 & 33.7068 & 15.536 & 0.003 & 161 & Kra & C \\
36.4837 & 36.6453 & 15.853 & 0.004 & 122 & UJH & C \\
36.5406 & 36.6977 & 15.912 & 0.002 & 137 & Kra & C \\
36.6880 & 36.8481 & 15.945 & 0.005 & 100 & GFB & C \\
37.2884 & 37.3456 & 16.198 & 0.005 & 76 & PXR & V \\
37.5412 & 37.6965 & 16.031 & 0.002 & 134 & Kra & C \\
37.6903 & 37.8487 & 16.088 & 0.005 & 100 & GFB & C \\
37.9127 & 38.1294 & 16.418 & 0.004 & 524 & KU & C \\
38.5368 & 38.6939 & 16.851 & 0.007 & 136 & Kra & C \\
38.9180 & 39.1222 & 18.341 & 0.021 & 500 & KU & C \\
39.5366 & 39.6909 & 18.923 & 0.017 & 103 & Kra & C \\
43.2612 & 43.2671 & 16.972 & 0.034 & 5 & deM & C \\
43.5344 & 43.5405 & 16.414 & 0.013 & 5 & Kra & C \\
43.7419 & 43.8332 & 16.395 & 0.005 & 58 & GFB & C \\
44.2618 & 44.3051 & 16.866 & 0.007 & 30 & deM & C \\
44.7044 & 44.8490 & 17.551 & 0.013 & 91 & GFB & C \\
45.2640 & 45.2699 & 18.184 & 0.045 & 5 & deM & C \\
45.6895 & 45.8433 & 18.768 & 0.033 & 86 & GFB & C \\
46.6888 & 46.8426 & 16.502 & 0.006 & 96 & GFB & C \\
50.2631 & 50.2720 & 17.189 & 0.016 & 7 & deM & C \\
51.4569 & 51.4569 & 17.100 & -- & 1 & MUY & C \\
54.6967 & 54.8100 & 18.469 & 0.098 & 48 & GFB & C \\
55.3672 & 55.3738 & 17.007 & 0.031 & 10 & PXR & V \\
55.6799 & 55.8201 & 16.600 & 0.013 & 85 & GFB & C \\
56.7061 & 56.8165 & 18.106 & 0.049 & 66 & GFB & C \\
57.6870 & 57.8128 & 19.348 & 0.033 & 78 & GFB & C \\
\hline
  \multicolumn{7}{l}{\commenta BJD$-$2456200.} \\
  \multicolumn{7}{l}{\commentb Mean magnitude.} \\
  \multicolumn{7}{l}{\commentc 1-$\sigma$ of the mean magnitude.} \\
  \multicolumn{7}{l}{\commentd Number of observations.} \\
  \multicolumn{7}{l}{\commente Observer's code.} \\
  \multicolumn{7}{l}{\commentf Filter.  ``C'' means no filter (clear).} \\
\end{tabular}
\end{center}
\end{table}

The photometry log is given in tables \ref{tab:log} and \ref{tab:log_j2037}.
The continuation of table \ref{tab:log} can be seen on the electronic version.
The observations are composed of those obtained by
 observers listed in table \ref{tab:obs}
 and AAVSO observers.
Comparison stars are shown in tables \ref{tab:comp} and 
\ref{tab:comp_j2037}. The unfilter CCD magnitudes are close to $V$
 for outbursting CVs.

All the observed times were corrected to barycentric Julian days
(BJDs).  The log of the observation is listed in table \ref{sec:obs}.  
Before making the analysis, we corrected zero-point differences
between different observers by adding a constant to each observer.

In making period analysis, we used phase dispersion minimization
(PDM) method \citep{PDM}.  We subtracted the global trend
of the outburst light curve by subtracting smoothed light
curve obtained by locally-weighted polynomial regression (LOWESS,
\cite{LOWESS}) before making the PDM analysis.
The 1-$\sigma$ error in periods of the PDM analysis was determined
by the methods in \citet{fer89error}, \citet{Pdot2}.

We used a variety of bootstrapping in
estimating the robustness of the result of PDM.
We typically analyzed 100 samples which randomly contain 50\% of
observations, and performed PDM analysis for these samples.
The bootstrap result is shown as a form of 90\% confidence intervals
in the resultant $\theta$ statistics.

\begin{table*}
\caption{Observers and equipment.}
\label{tab:obs}
\begin{center}
\begin{tabular}{ccc}
\hline
Code \commenta & Observer/Site & Equipment \\
\hline
Mas & G. Masi &   43.2 cm telescope and STL6303E \\
deM & E. de Miguel &  28cm telescope and QSI-518 wsg \\
UJH & J. Ulowetz & 23.5cm telescope and QSI-583 wsg \\
LCO &  C. Littlefield &  - \\
KU & Kyoto University & 40cm telescope and ST-9 CCD \\
GFB & W. Goff & - \\
Kra & T. Krajci & 28cm telescope \\
Mhh & H. Maehara &  25cm telescope and ST-7XME \\
SWI & W. Stein & 35.56 cm telescope and ST10-XME \\
SRI & R. Sabo  & - \\
OKU & Osaka Kyoiku U. & 51cm telescope and ST-10 CCD \\
DPV & P. Dubovsky & 28 cm telescope and Meade DSI ProII
CCD \\
 & & 35.6 cm and Moravian Instruments G2-1600 CCD \\
DKS & S. Dvorak & - \\
HMB & F.--J. Hambsch & 40cm telescope and FLI ML16803 CCD \\
PXR & R. Pickard & - \\
MEV & E. Morelle & - \\
PSD & S. Padovan & - \\
MUY & E. Muyllaert & - \\
\hline
\multicolumn{3}{l}{\commenta  Observer's code used in table \ref{tab:log} and
 table \ref{tab:log_j2037}.} \\
\end{tabular}
\end{center}
\end{table*}

\begin{table}
\caption{Comparison stars for MASTER J211258.}\label{tab:comp}
\begin{center}
\begin{tabular}{cc}
\hline
Observers & Comparison star\\
\hline
deM HMB & GSC 2186.232 \\
Mhh & GSC 2190.396 \\
OKU SWI DPV UJH & GSC 2190.309 \\
OKU & GSC 2190.546 \\
\hline
\end{tabular}
\end{center}
\end{table}

\begin{table}
\caption{Comparison stars for MASTER J203749.}\label{tab:comp_j2037}
\begin{center}
\begin{tabular}{cc}
\hline
Observers & Comparison star\\
\hline
deM Kra  & GSC 3954.1468 \\
 KU & GSC 3954.649 \\
Mhh & GSC 3954.2015 \\
\hline

\end{tabular}
\end{center}
\end{table}

\section{MASTER OT J211258.65+242145.4}\label{sec:res2112}

\subsection{Overall Light Curve}\label{sec:overall2112}

\begin{figure}
  \begin{center}
    \FigureFile(88mm,110mm){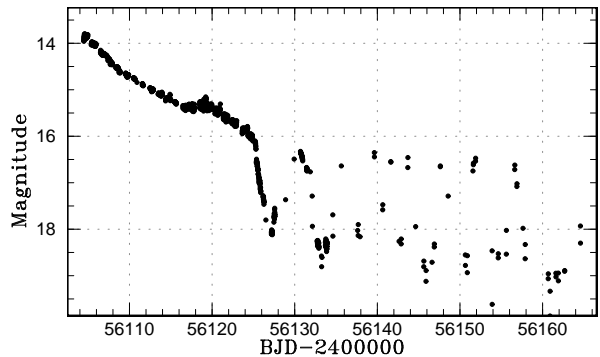}
  \end{center}
  \caption{Overall light curve of MASTER J211258. The data were binned to 0.01 d.}
  \label{fig:lightcurve}
\end{figure}

   Figure \ref{fig:lightcurve} shows the overall light curve
of MASTER J211258.
The object was first detected in superoutburst on June 24
 (BJD 2456104).
Although the early rise was hardly observed,
 the recorded possible maximum brightness of $V$ = 13.72
 on about BJD 2456104.57 and the long duration of existence of early
 superhumps (subsection \ref{sec:earlysh2112}) suggest that the
 detection was made within a few days
 of the start of the superoutburst.
The main superoutburst lasted until BJD 2456125, followed
 by a rapid decline. The object then faded below $V$ = 18.
 On BJD 2456130, the first rebrightening ($V$ = 16.32 at
 maximum) was recorded.
 Similar rebrightenings repetitively occured eight times in total.

The light curve is composed of different segments:
 BJD 2456104--2456116, during which early superhumps were present,
 BJD 2456116--2456126, during which superhumps were present,
 and the following eight rebrightenings starting on BJD 2456130.

\subsection{Early Superhumps}\label{sec:earlysh2112}

 Since the orbital period is considered to be very close to the 
early superhump period \citep{ish02rzleoproc}, we used
 the period of the early superhump
 as the orbital period.

\begin{figure}
  \begin{center}
    \FigureFile(88mm,110mm){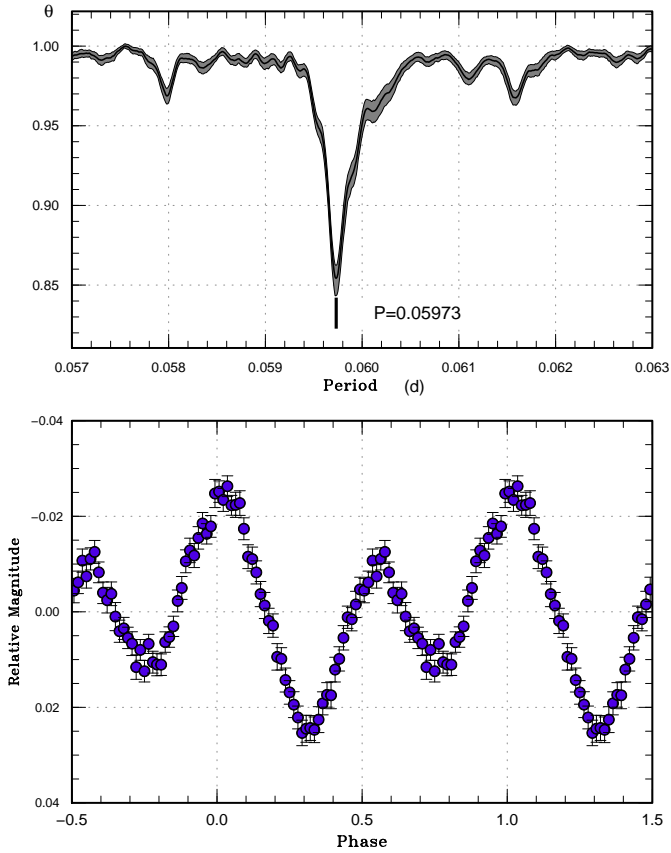}
  \end{center}
  \caption{Early superhumps in MASTER J211258 (BJD 2456104--2456116). (Upper): PDM analysis.
     (Lower): Phase-averaged profile.}
  \label{fig:j2112eshpdm}
\end{figure}

\begin{figure}
  \begin{center}
    \FigureFile(88mm,110mm){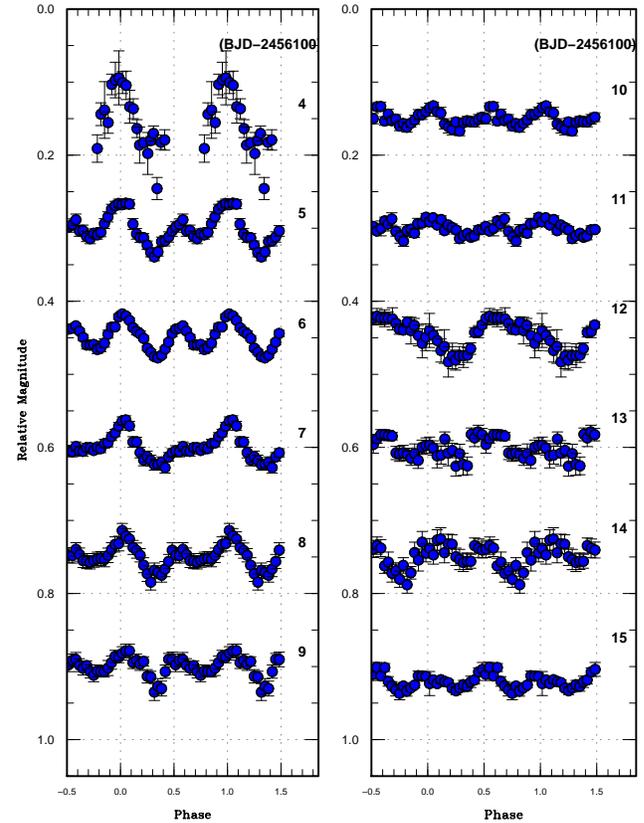}
  \end{center}
  \caption{Nightly variation of the profile of early superhumps in MASTER J211258.}
  \label{fig:pro_early}
\end{figure}

The early superhumps were fortunately well recorded.
 The doubly humped early superhumps with a period of 0.059732(3)~d
 were recorded (figure \ref{fig:j2112eshpdm}) during the earliest
 12 nights of observations. 
The mean amplitude of early superhumps was $\sim$0.05 mag.
Figure \ref{fig:pro_early} shows the nightly variation of the profile
of early superhumps.

\subsection{Ordinary Superhumps}\label{sec:ordinarysh2112}

The fractional superhump excess, an observational measure of
the precession rate of the accretion disk, is defined by $\epsilon$,
 as mentioned in the introduction.

\begin{figure}
  \begin{center}
    \FigureFile(88mm,110mm){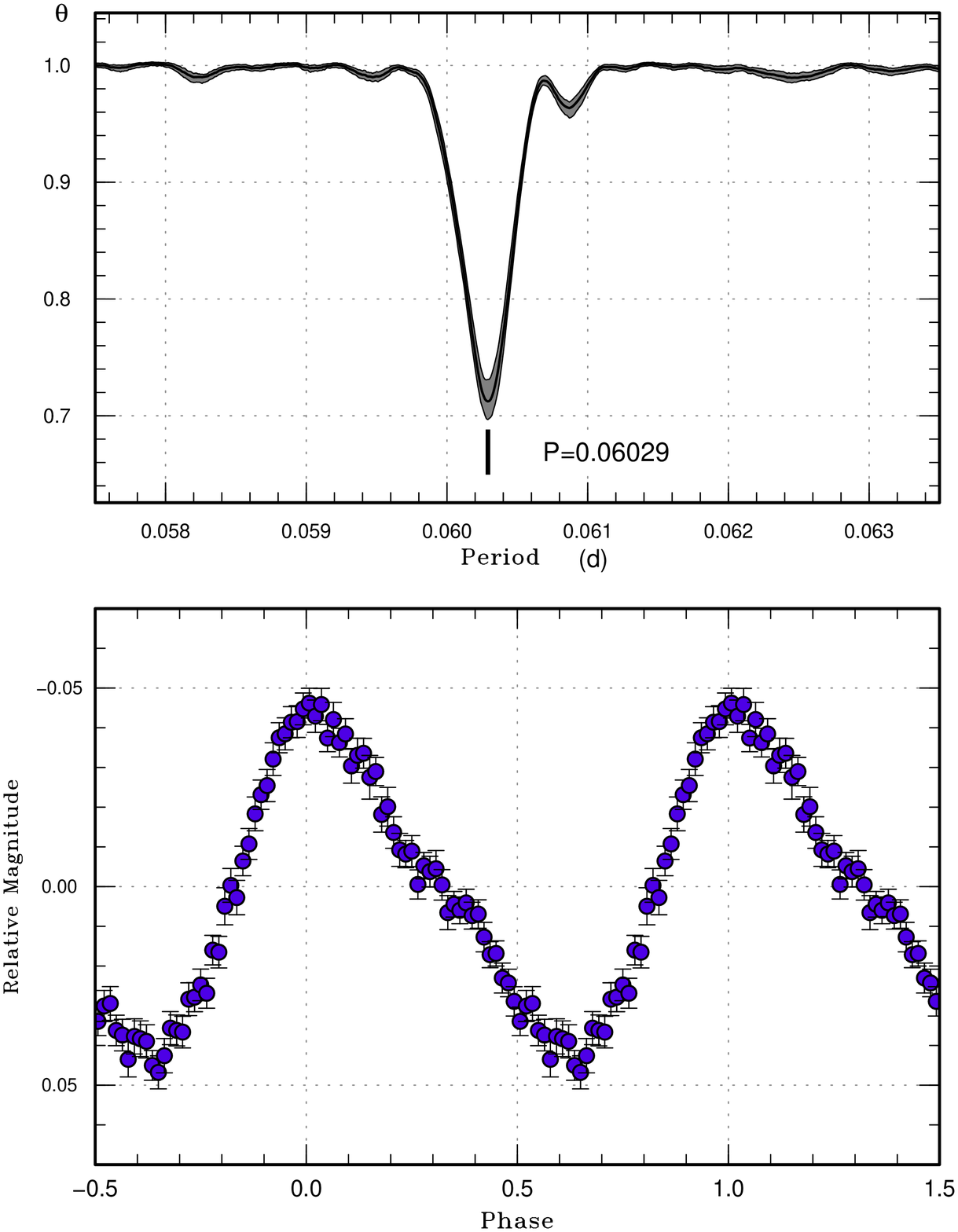}
  \end{center}
  \caption{Ordinary superhumps in MASTER J211258 (BJD 2456116--2456126). (Upper): PDM analysis.
     (Lower): Phase-averaged profile.}
  \label{fig:j2112shpdm}
\end{figure}

\begin{figure}
  \begin{center}
    \FigureFile(88mm,110mm){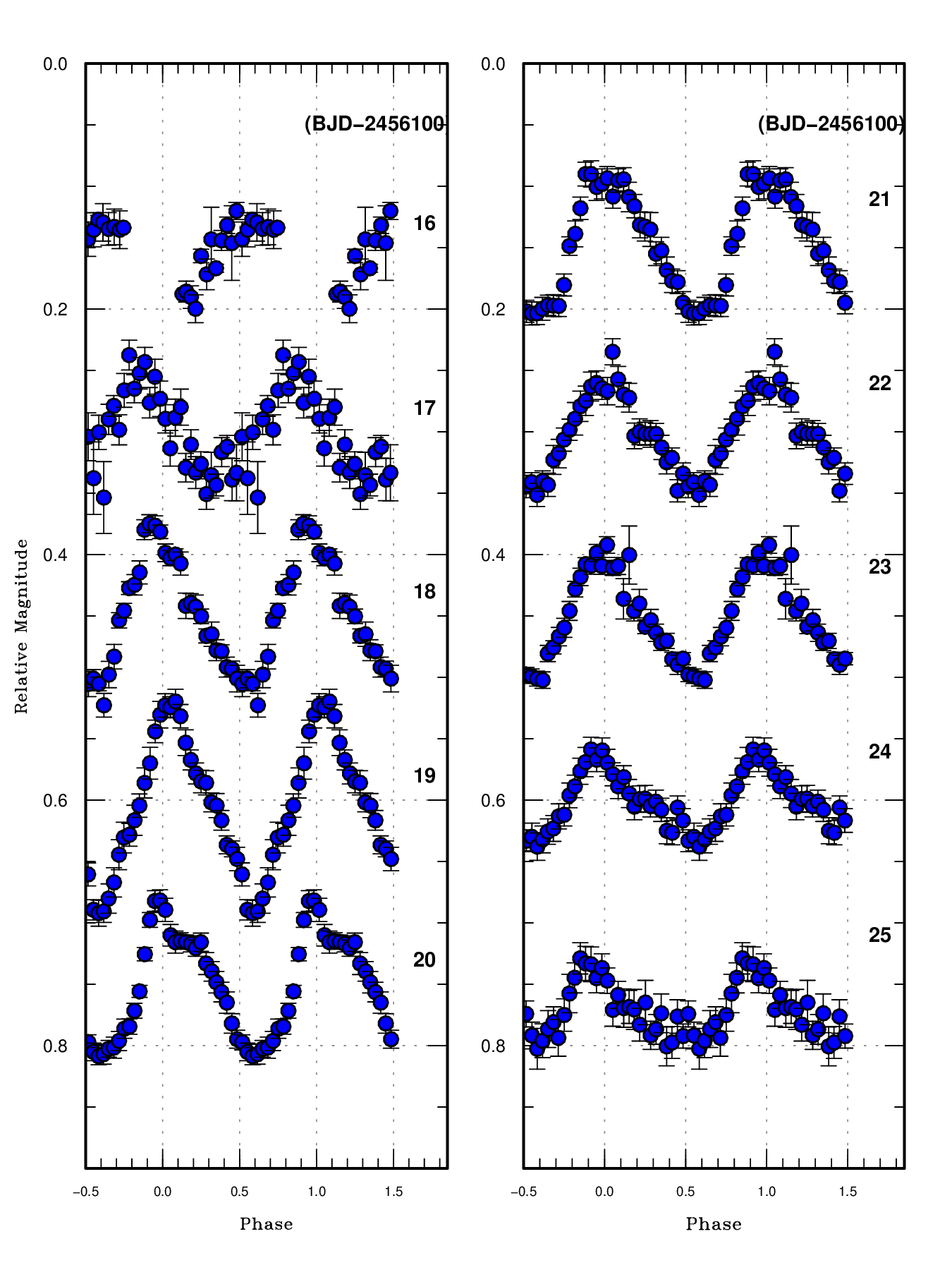}
  \end{center}
  \caption{Nightly variation of the profile of ordinary superhumps in MASTER J211258.}
  \label{fig:pro_ordinary}
\end{figure}

A period analysis indicated the presence of a very stable period
 of 0.060291(4) d during  BJD 2456116--2456126 (figure \ref{fig:j2112shpdm}).
Figure \ref{fig:pro_ordinary} shows the nightly variation of the
 profile of ordinary superhumps.
The amplitude of superhumps was 0.1$-$0.2 mag.
They became larger until BJD 2456119, and then became smaller.
  The fractional superhump excess was 0.009376(7).

\subsection{$O-C$ Analysis of Ordinary Superhumps}\label{sec:ocanalysis2112}

   We determined the times of maxima of ordinary superhumps
 in the way given in \citet{Pdot}.  The resultant times
are listed in table \ref{tab:j2112sh}.

\begin{figure}
  \begin{center}
    \FigureFile(88mm,110mm){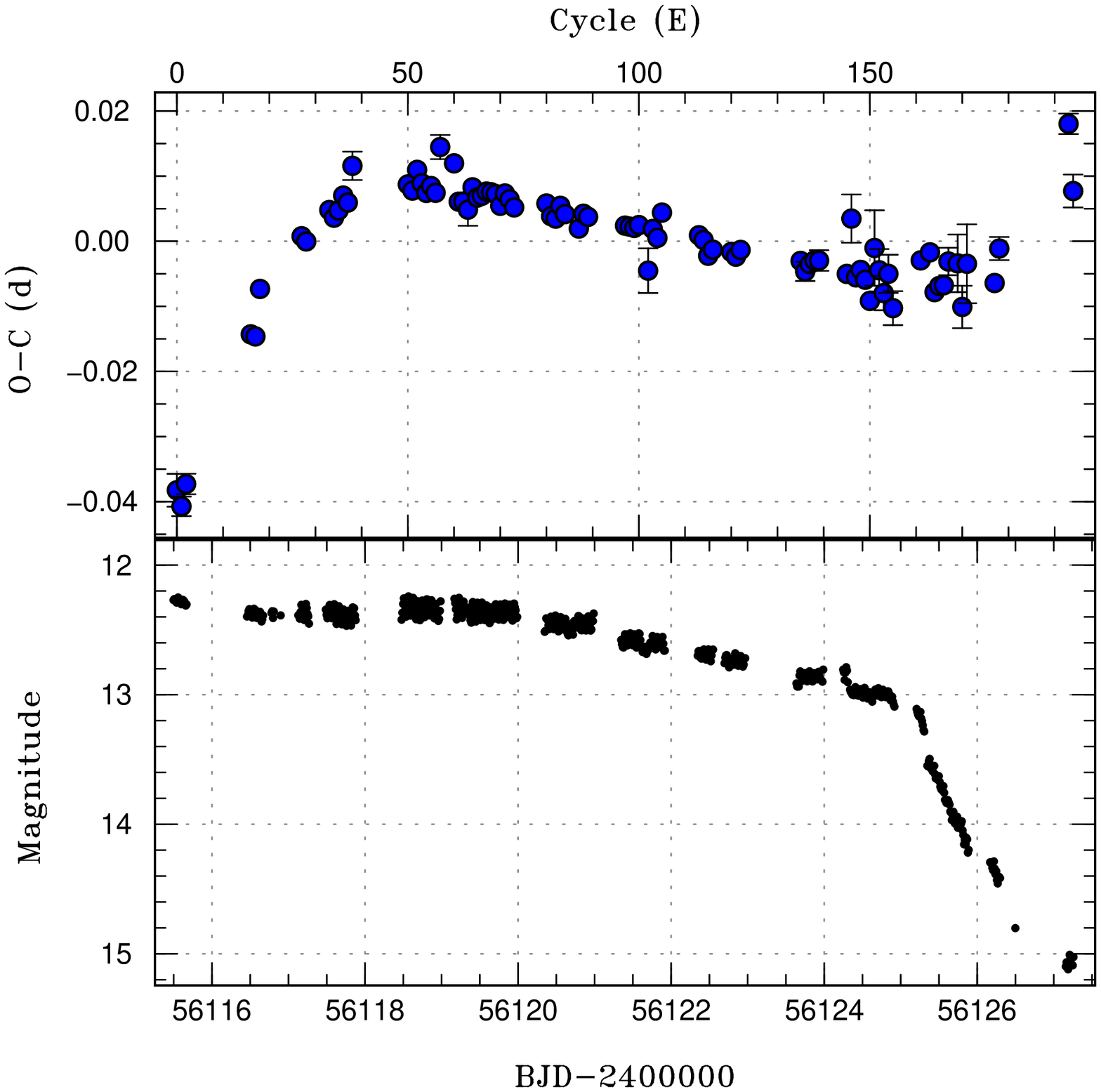}
  \end{center}
  \caption{The $O-C$ curve of MASTER J211258 during the superoutburst. 
  (Upper:) $O-C$ diagram of superhumps in MASTER J211258.
   An ephemeris of BJD 2456115.541+0.060375$E$ was used to draw this
   figure. (Lower:) Light curve. The data were binned to 0.01 d.}
  \label{fig:ocplot}
\end{figure}

The $O-C$ curve of MASTER J2112 is shown in figure \ref{fig:ocplot}.
The stage A ($E \le 38$) and stage B ($E \ge 50$) are clearly seen.
 In stage A, superhumps have a long period and the period is decreasing.
 In stage B, $P_{\rm dot}$ (=$\dot{P}/P$) is almost zero.
The mean periods for these stages were  0.06158(5) d (stage A) and
 0.060221(9) d (stage B), respectively.

The $P_{\rm dot}$ for stage B ($E \ge 50$) was $+0.8(1.0) \times 10^{-5}$.
The peak of the superhump amplitudes is close to the time of stage A-B
 transition.
A major increase in the period took place during the final part
 of stage B ($E \ge 50$). 

 During stage A, superhumps with a period of 0.06158(5) d
 were observed and the
fractional superhump excess was 0.0310(1).
 
\begin{table}
\caption{Times of superhump maxima in MASTER J211258.}\label{tab:j2112sh}
\begin{center}
\begin{tabular}{ccccc}
\hline
$E$ & maximum time\commenta & error & $O-C$\commentb & $N$\commentc \\
\hline
0 & 56115.5028 & 0.0025 & $-$0.0382 & 66 \\
1 & 56115.5607 & 0.0015 & $-$0.0407 & 67 \\
2 & 56115.6245 & 0.0016 & $-$0.0372 & 59 \\
16 & 56116.4927 & 0.0014 & $-$0.0143 & 66 \\
17 & 56116.5528 & 0.0012 & $-$0.0146 & 66 \\
18 & 56116.6204 & 0.0013 & $-$0.0073 & 63 \\
27 & 56117.1719 & 0.0012 & 0.0008 & 105 \\
28 & 56117.2315 & 0.0011 & 0.0000 & 125 \\
33 & 56117.5382 & 0.0005 & 0.0048 & 66 \\
34 & 56117.5974 & 0.0005 & 0.0037 & 60 \\
35 & 56117.6588 & 0.0005 & 0.0048 & 76 \\
36 & 56117.7215 & 0.0009 & 0.0071 & 24 \\
37 & 56117.7808 & 0.0010 & 0.0060 & 26 \\
38 & 56117.8468 & 0.0022 & 0.0116 & 18 \\
50 & 56118.5685 & 0.0003 & 0.0088 & 67 \\
51 & 56118.6279 & 0.0004 & 0.0078 & 71 \\
52 & 56118.6915 & 0.0012 & 0.0110 & 18 \\
53 & 56118.7498 & 0.0006 & 0.0090 & 44 \\
54 & 56118.8087 & 0.0005 & 0.0075 & 70 \\
55 & 56118.8701 & 0.0006 & 0.0085 & 67 \\
56 & 56118.9294 & 0.0008 & 0.0075 & 40 \\
57 & 56118.9968 & 0.0019 & 0.0145 & 23 \\
60 & 56119.1755 & 0.0013 & 0.0120 & 46 \\
61 & 56119.2299 & 0.0007 & 0.0061 & 97 \\
62 & 56119.2903 & 0.0013 & 0.0061 & 47 \\
63 & 56119.3495 & 0.0025 & 0.0049 & 36 \\
64 & 56119.4133 & 0.0003 & 0.0083 & 63 \\
65 & 56119.4721 & 0.0004 & 0.0068 & 86 \\
66 & 56119.5327 & 0.0004 & 0.0070 & 117 \\
67 & 56119.5938 & 0.0005 & 0.0077 & 45 \\
68 & 56119.6540 & 0.0005 & 0.0076 & 75 \\
69 & 56119.7141 & 0.0011 & 0.0073 & 29 \\
70 & 56119.7727 & 0.0005 & 0.0055 & 80 \\
71 & 56119.8350 & 0.0006 & 0.0074 & 81 \\
72 & 56119.8945 & 0.0007 & 0.0065 & 75 \\
73 & 56119.9536 & 0.0006 & 0.0053 & 62 \\
80 & 56120.3768 & 0.0006 & 0.0059 & 62 \\
81 & 56120.4352 & 0.0005 & 0.0039 & 63 \\
82 & 56120.4952 & 0.0004 & 0.0035 & 111 \\
83 & 56120.5576 & 0.0005 & 0.0055 & 124 \\
84 & 56120.6167 & 0.0007 & 0.0042 & 75 \\
87 & 56120.7956 & 0.0006 & 0.0020 & 128 \\
88 & 56120.8582 & 0.0008 & 0.0043 & 110 \\
89 & 56120.9181 & 0.0007 & 0.0038 & 66 \\
97 & 56121.3998 & 0.0005 & 0.0025 & 64 \\
98 & 56121.4600 & 0.0005 & 0.0023 & 64 \\
99 & 56121.5202 & 0.0005 & 0.0021 & 63 \\
100 & 56121.5810 & 0.0004 & 0.0026 & 62 \\
102 & 56121.6947 & 0.0034 & $-$0.0044 & 16 \\
103 & 56121.7615 & 0.0010 & 0.0020 & 80 \\
104 & 56121.8205 & 0.0008 & 0.0006 & 129 \\
105 & 56121.8848 & 0.0007 & 0.0045 & 132 \\
\hline
  \multicolumn{5}{l}{\commenta BJD$-$2400000.} \\
  \multicolumn{5}{l}{\commentb  $C= 2456115.5410 + 0.060375 E$.} \\
  \multicolumn{5}{l}{\commentc Number of points used to determine the maximum.} \\
\end{tabular}
\end{center}
\end{table}

\addtocounter{table}{-1}
\begin{table}
\caption{Times of superhump maxima in MASTER J211258 (continued).}
\begin{center}
\begin{tabular}{ccccc}
\hline
$E$ & maximum time\commenta & error & $O-C$\commentb & $N$\commentc \\
\hline
113 & 56122.3643 & 0.0008 & 0.0010 & 49 \\
114 & 56122.4239 & 0.0005 & 0.0003 & 64 \\
115 & 56122.4819 & 0.0004 & $-$0.0022 & 63 \\
116 & 56122.5432 & 0.0007 & $-$0.0012 & 44 \\
120 & 56122.7844 & 0.0007 & $-$0.0016 & 146 \\
121 & 56122.8440 & 0.0007 & $-$0.0023 & 148 \\
122 & 56122.9054 & 0.0006 & $-$0.0013 & 136 \\
135 & 56123.6886 & 0.0010 & $-$0.0029 & 83 \\
136 & 56123.7474 & 0.0015 & $-$0.0046 & 137 \\
137 & 56123.8089 & 0.0009 & $-$0.0034 & 169 \\
138 & 56123.8698 & 0.0011 & $-$0.0029 & 175 \\
139 & 56123.9302 & 0.0016 & $-$0.0029 & 66 \\
145 & 56124.2904 & 0.0010 & $-$0.0049 & 92 \\
146 & 56124.3592 & 0.0037 & 0.0036 & 43 \\
147 & 56124.4106 & 0.0009 & $-$0.0055 & 52 \\
148 & 56124.4721 & 0.0011 & $-$0.0043 & 117 \\
149 & 56124.5310 & 0.0007 & $-$0.0058 & 102 \\
150 & 56124.5881 & 0.0013 & $-$0.0091 & 58 \\
151 & 56124.6566 & 0.0058 & $-$0.0009 & 58 \\
152 & 56124.7135 & 0.0033 & $-$0.0044 & 64 \\
153 & 56124.7704 & 0.0026 & $-$0.0079 & 64 \\
154 & 56124.8337 & 0.0029 & $-$0.0049 & 65 \\
155 & 56124.8888 & 0.0026 & $-$0.0102 & 72 \\
161 & 56125.2584 & 0.0007 & $-$0.0028 & 66 \\
163 & 56125.3804 & 0.0010 & $-$0.0016 & 43 \\
164 & 56125.4347 & 0.0010 & $-$0.0077 & 59 \\
165 & 56125.4960 & 0.0014 & $-$0.0068 & 88 \\
166 & 56125.5565 & 0.0012 & $-$0.0067 & 83 \\
167 & 56125.6205 & 0.0021 & $-$0.0030 & 65 \\
169 & 56125.7410 & 0.0044 & $-$0.0033 & 20 \\
170 & 56125.7947 & 0.0033 & $-$0.0100 & 29 \\
171 & 56125.8617 & 0.0061 & $-$0.0034 & 23 \\
177 & 56126.2210 & 0.0012 & $-$0.0063 & 66 \\
178 & 56126.2866 & 0.0018 & $-$0.0010 & 49 \\
193 & 56127.2114 & 0.0016 & 0.0181 & 66 \\
194 & 56127.2615 & 0.0025 & 0.0078 & 66 \\
\hline
  \multicolumn{5}{l}{\commenta BJD$-$2400000.} \\
  \multicolumn{5}{l}{\commentb $C= 2456115.5410 + 0.060375 E$.} \\
  \multicolumn{5}{l}{\commentc Number of points used to determine the maximum.} \\
\end{tabular}
\end{center}
\end{table}

\section{MASTER OT J203749.39+552210.3}\label{sec:res2037}

\subsection{Overall Light Curve}\label{sec:overall2037}
\begin{figure}
  \begin{center}
    \FigureFile(88mm,110mm){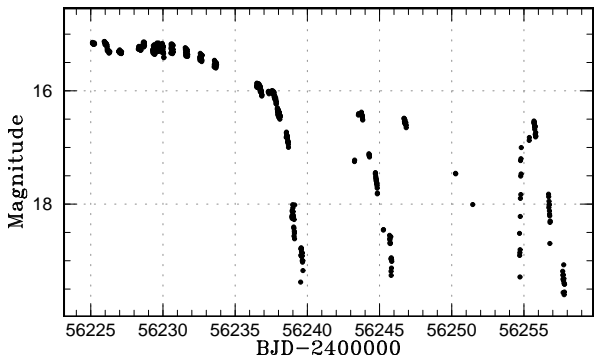}
  \end{center}
  \caption{Overall light curve of MASTER J203749. The data were binned to 0.01 d.}
  \label{fig:lightcurve_j2037}
\end{figure}

 Figure \ref{fig:lightcurve_j2037} shows the overall light curve
of MASTER J203749. 
The object was first detected in superoutburst
 ($V$ = 15.1 at maximum) on October 22
 (BJD 2456225).
The short duration of the recorded early superhump stage
 suggests that the earlier part of the outburst was missed.
The main superoutburst lasted until BJD 2456238, followed
 by a rapid decline. The object then faded below $V$ = 19.
 On BJD 2456243, the first rebrightening ($V$ = 16.37 at
 maximum) was recorded.
 Similar rebrightenings repetitively occured at least
 four times in total, although the number of observations
 was not sufficient to detect other potential rebrightenings.

The object developed early superhumps during BJD 2456225--2456227.
 During BJD 2456227--2456238, superhumps appeared.

\subsection{Early Superhumps}\label{sec:earlysh2037}

\begin{figure}
  \begin{center}
    \FigureFile(88mm,110mm){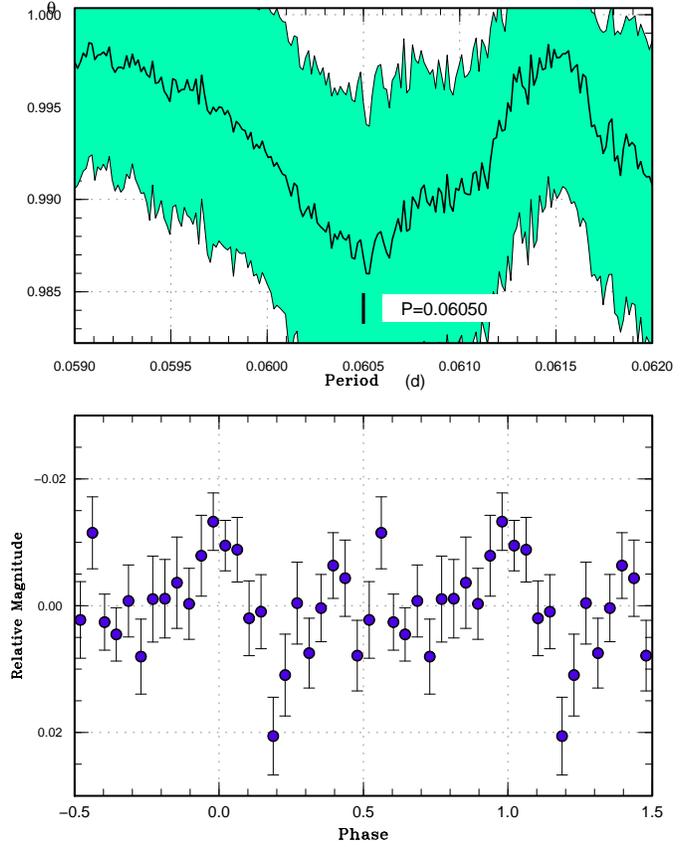}
  \end{center}
  \caption{Early superhumps in MASTER J203749 (BJD 2456225--2456227). (Upper): PDM analysis.
     (Lower): Phase-averaged profile.}
  \label{fig:j2037eshpdm}
\end{figure}

\begin{figure}
  \begin{center}
    \FigureFile(70mm,90mm){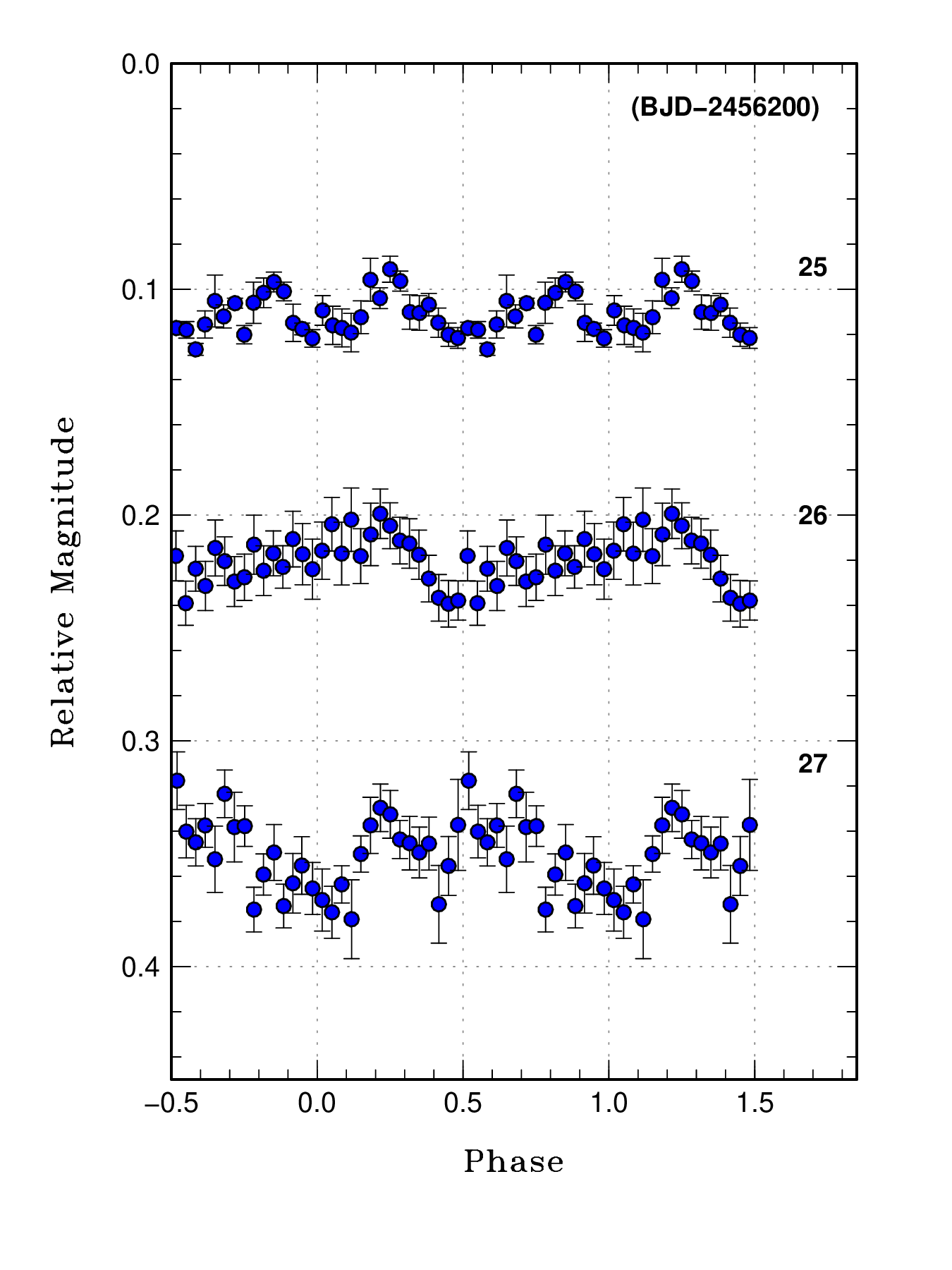}
  \end{center}
  \caption{Nightly variation of the profile of early superhumps in MASTER J203749.} 
  \label{fig:pro_early_j2037}
\end{figure}

It was likely that the final stage of early superhumps was
 observed. The mean period of 0.06051(18) d
 was recorded during BJD 2456225--2456227 (figure \ref{fig:j2037eshpdm}).
We identified this period to be the orbital period.
Figure \ref{fig:pro_early_j2037} shows the nightly variation of the profile
of early superhumps.
The data give a larger error than in MASTER J211258 because
 a period of the observations was short than in MASTER J211258.

\subsection{Ordinary Superhumps}\label{sec:ordinarysh2037}

\begin{figure}
  \begin{center}
    \FigureFile(88mm,110mm){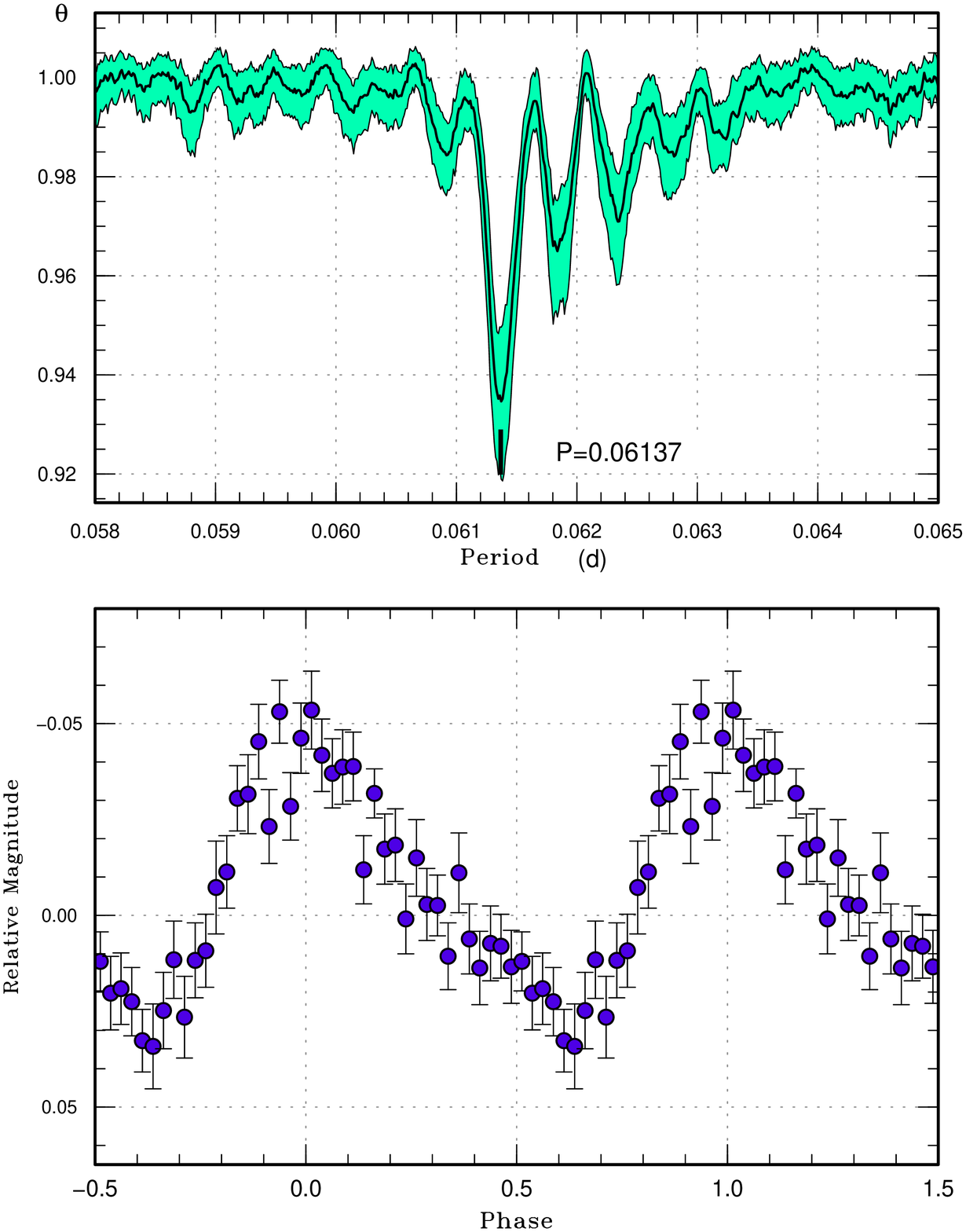}
  \end{center}
  \caption{Ordinary superhumps in MASTER J203749 (BJD 2456227--2456238). (Upper): PDM analysis.
     (Lower): Phase-averaged profile.}
  \label{fig:j2037shpdm}
\end{figure}

\begin{figure}
  \begin{center}
    \FigureFile(88mm,110mm){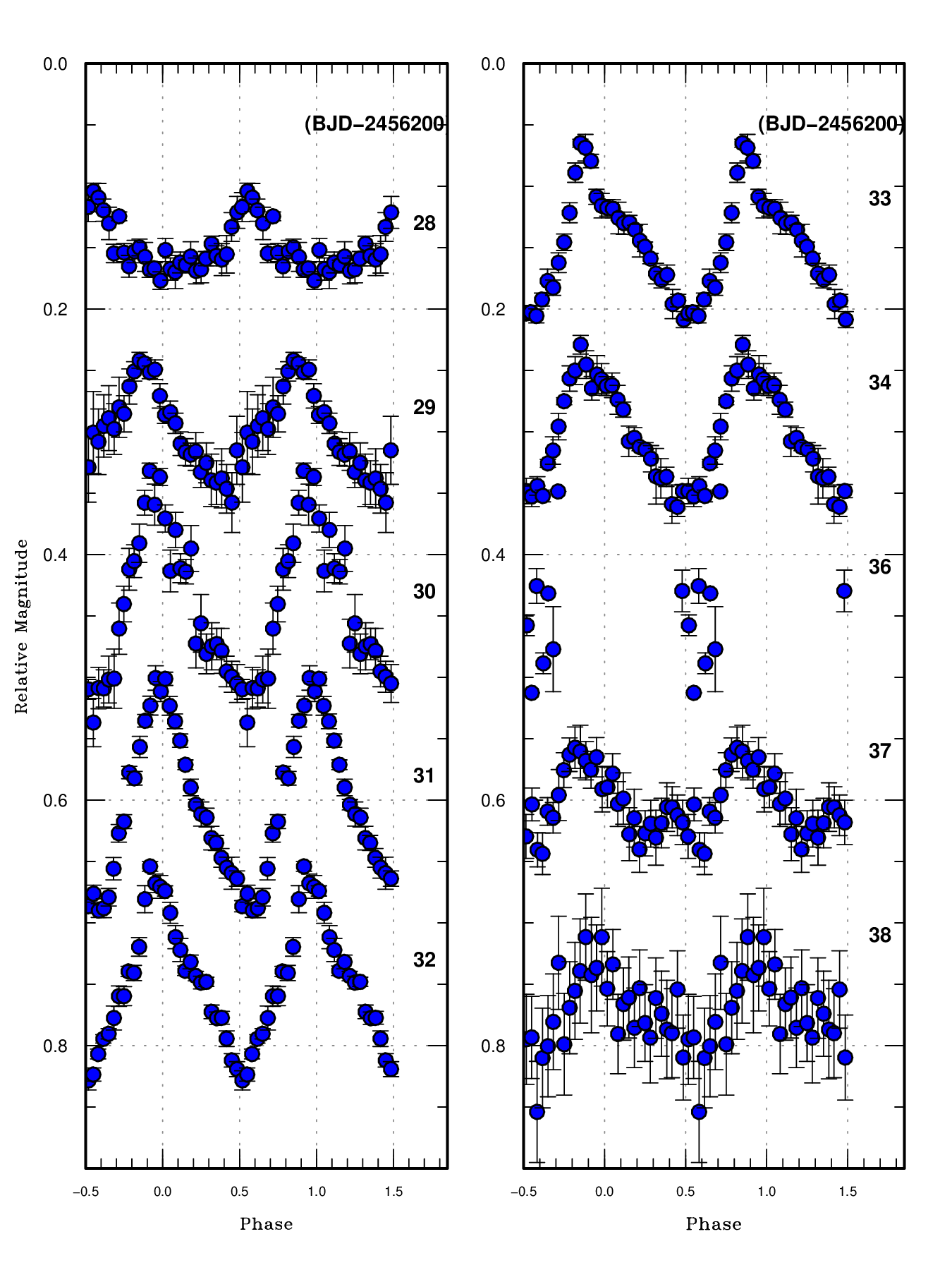}
  \end{center}
  \caption{Nightly variation of the profile of ordinary superhumps in MASTER J203749.}
  \label{fig:pro_ordinary_j2037}
\end{figure}

Ordinary superhumps with a period of 0.061307(9) d developed
  during BJD 2456227--2456238 (figure \ref{fig:j2037shpdm}).
Figure \ref{fig:pro_ordinary_j2037} shows the nightly variation of the
 profile of ordinary superhumps.
 The growth of ordinary superhumps was clearly seen.
 The amplitude increased until around BJD 2456231, and then became
 smaller.
 The fractional superhump excess was 0.01334(15).

\subsection{$O-C$ Analysis of Ordinary Superhumps}\label{sec:ocanalysis2037}

\begin{figure}
  \begin{center}
    \FigureFile(88mm,110mm){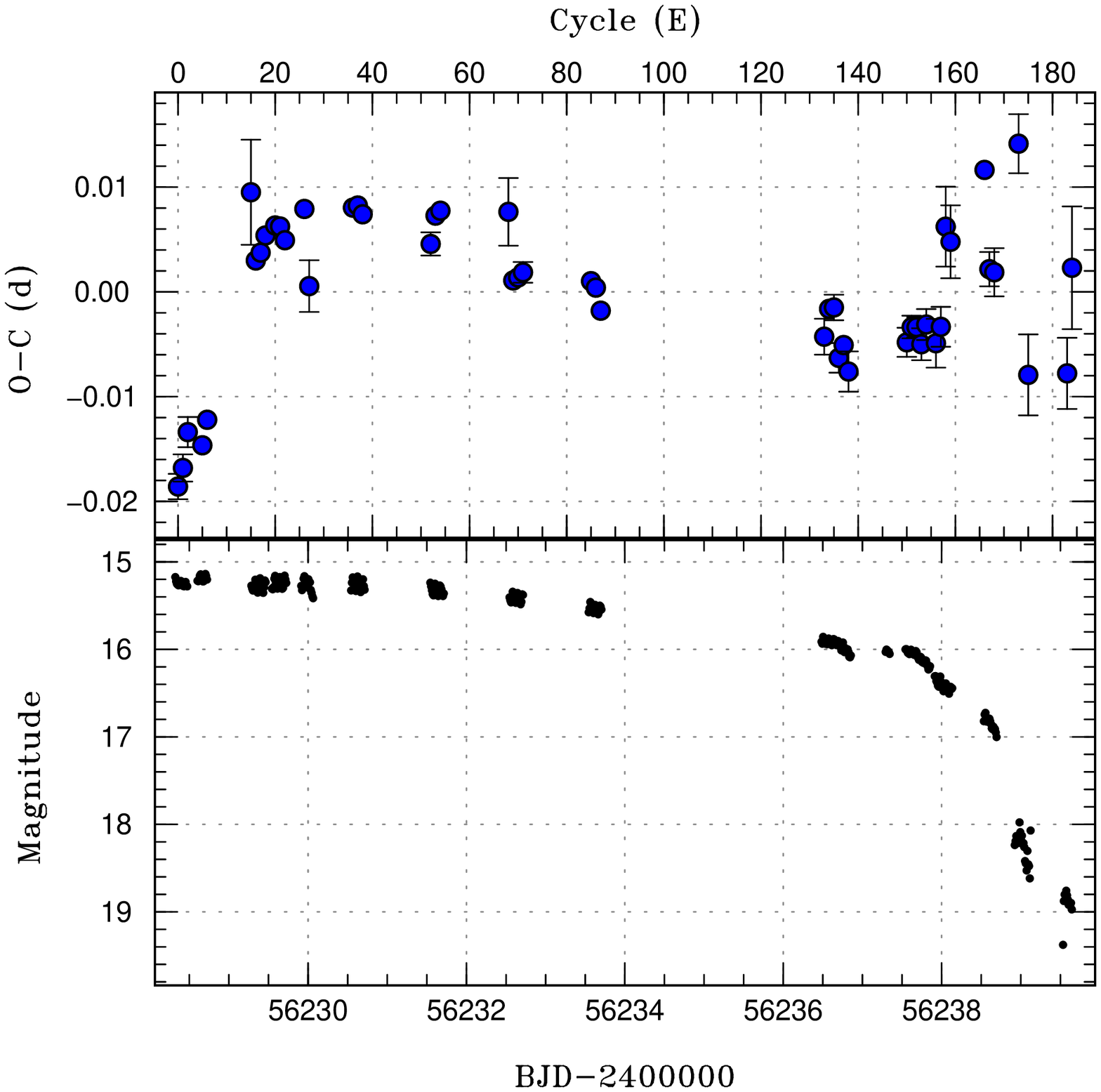}
  \end{center}
  \caption{The $O-C$ curve of MASTER J203749 during the superoutburst. 
  (Upper:) $O-C$ diagram of superhumps in MASTER J203749.
   An ephemeris of BJD 2456228.358+0.060135$E$ was used to draw this
   figure. (Lower:) Light curve. The data were binned to 0.01 d.}
  \label{fig:ocplot_j2037}
\end{figure}

  The times of superhump maxima
are listed in table \ref{tab:j2037sh}.
The $O-C$ curve of MASTER OT J2037 is shown in figure \ref{fig:ocplot_j2037},
 clearly composed of stages A ($E \le 30$) and B ($E \ge 36$).
The mean periods for these stages were  0.06271(11) d (stage A) and
 0.061307(9) d (stage B), respectively.

Disregarding stage A superhumps ($E \le 30$), a marginally
 positive $P_{\rm dot}$ of $+2.9(1.0) \times 10^{-5}$ was recorded.
A major increase in the period also took place during the final part
 of stage B ($E \ge 36$).  

During stage A, superhumps with a period of 0.06271(11) d
 were recorded and the
 fractional superhump excess was 0.0365(2).

The analysis showed $P_{\rm dot}$ during stage B of both MASTER
 J211258 and MASTER J203749 were positive.
The positive period derivatives were reported in some dwarf novae
 including WZ Sge-type dwarf novae.
 The mechanism that causes positive period derivatives of superhump periods
is unknown.

\begin{table}
\caption{Times of superhump maxima in MASTER J203749.}\label{tab:j2037sh}
\begin{center}
\begin{tabular}{ccccc}
\hline
$E$ & maximum time\commenta & error & $O-C$\commentb & $N$\commentc \\
\hline
0 & 56228.33 & 0.0012 & $-$0.0186 & 62 \\
1 & 56228.39 & 0.0013 & $-$0.0168 & 62 \\
2 & 56228.46 & 0.0014 & $-$0.0134 & 54 \\
5 & 56228.64 & 0.0006 & $-$0.0146 & 60 \\
6 & 56228.70 & 0.0007 & $-$0.0122 & 51 \\
15 & 56229.28 & 0.0050 & 0.0095 & 23 \\
16 & 56229.33 & 0.0006 & 0.0030 & 59 \\
17 & 56229.39 & 0.0005 & 0.0037 & 62 \\
18 & 56229.46 & 0.0006 & 0.0054 & 44 \\
20 & 56229.58 & 0.0003 & 0.0063 & 47 \\
21 & 56229.64 & 0.0003 & 0.0062 & 47 \\
22 & 56229.70 & 0.0006 & 0.0049 & 37 \\
26 & 56229.95 & 0.0006 & 0.0079 & 125 \\
27 & 56230.01 & 0.0025 & 0.0005 & 75 \\
36 & 56230.57 & 0.0003 & 0.0080 & 46 \\
37 & 56230.63 & 0.0003 & 0.0082 & 47 \\
38 & 56230.69 & 0.0003 & 0.0074 & 47 \\
52 & 56231.55 & 0.0011 & 0.0046 & 27 \\
53 & 56231.61 & 0.0004 & 0.0073 & 47 \\
54 & 56231.67 & 0.0006 & 0.0077 & 47 \\
68 & 56232.53 & 0.0032 & 0.0076 & 15 \\
69 & 56232.59 & 0.0005 & 0.0011 & 47 \\
70 & 56232.65 & 0.0006 & 0.0014 & 47 \\
71 & 56232.71 & 0.0010 & 0.0019 & 26 \\
85 & 56233.57 & 0.0004 & 0.0010 & 47 \\
86 & 56233.63 & 0.0005 & 0.0004 & 47 \\
87 & 56233.69 & 0.0007 & $-$0.0018 & 41 \\
133 & 56236.51 & 0.0017 & $-$0.0043 & 35 \\
134 & 56236.57 & 0.0009 & $-$0.0016 & 80 \\
135 & 56236.64 & 0.0012 & $-$0.0015 & 70 \\
136 & 56236.69 & 0.0014 & $-$0.0063 & 46 \\
137 & 56236.76 & 0.0008 & $-$0.0051 & 30 \\
138 & 56236.81 & 0.0019 & $-$0.0076 & 32 \\
150 & 56237.55 & 0.0014 & $-$0.0048 & 32 \\
151 & 56237.62 & 0.0011 & $-$0.0034 & 42 \\
152 & 56237.68 & 0.0010 & $-$0.0034 & 46 \\
153 & 56237.74 & 0.0015 & $-$0.0050 & 29 \\
154 & 56237.80 & 0.0015 & $-$0.0031 & 32 \\
156 & 56237.92 & 0.0023 & $-$0.0049 & 82 \\
157 & 56237.99 & 0.0019 & $-$0.0033 & 104 \\
158 & 56238.06 & 0.0038 & 0.0062 & 118 \\
159 & 56238.12 & 0.0035 & 0.0048 & 95 \\
166 & 56238.55 & 0.0009 & 0.0116 & 34 \\
167 & 56238.61 & 0.0016 & 0.0022 & 43 \\
168 & 56238.67 & 0.0023 & 0.0019 & 42 \\
173 & 56238.99 & 0.0028 & 0.0141 & 121 \\
175 & 56239.09 & 0.0039 & $-$0.0079 & 121 \\
183 & 56239.58 & 0.0034 & $-$0.0078 & 32 \\
184 & 56239.65 & 0.0059 & 0.0023 & 32 \\

\hline
  \multicolumn{5}{l}{\commenta BJD$-$2400000.} \\
  \multicolumn{5}{l}{\commentb  $C= 2456228.358+0.060135E$.} \\
  \multicolumn{5}{l}{\commentc Number of points used to determine the maximum.} \\
\end{tabular}
\end{center}
\end{table}

\section{Discussion}\label{sec:discussion}

\subsection{Brightening Associated with Evolution of Ordinary
Superhumps}\label{sec:brightening}
 In MASTER J211258, a sudden increase of the system brightness
 took place during the
 evolutionary stage of ordinary superhumps (BJD 2456116--2456118).
 In MASTER J203749, a possible increase of brightening was
 seen around BJD 2456229.5--2456230.

 This phenomenon is naturally explained by the increase of
 dissipation rate as a result of the growth of tidal instability.

 The systems with an increase of brightening during the
 evolutionary stage of ordinary superhumps are listed in table
 \ref{tab:brightening}.
 We can see a possible tendency that these systems exhibit
 long rebrightening or multiple rebrightenings.

\begin{table*}
\caption{The systems with the increase of brightening during
 evolutionary stage of ordinary superhumps.}\label{tab:brightening}
\begin{center}
\begin{tabular}{cccc}
\hline
Object & Year & Rebrightening\commenta & Reference \\
\hline
V466 And & 2008 & A & \citet{Pdot} \\
NN Cam\commentb & 2009 & - & \citet{Pdot2} \\
EG Cnc & 1996 & B & \citet{pat98egcnc} \\
AL Com & 1995 & A & \citet{nog97alcom}, \citet{pat96alcom}, \citet{Pdot} \\
V592 Her\commentb & 2010 & - & \citet{Pdot} \\
EZ Lyn & 2010 & B & \citet{Pdot3} \\
WZ Sge & 2001 & A & \citet{ish02wzsgeproc}, \citet{Pdot} \\
\hline
  \multicolumn{4}{l}{\commenta Classification of the rebrightenings
 of WZ Sge-type stars (subsection \ref{sec:rebrightenings}).} \\
 \multicolumn{4}{l}{\commentb Possible increase.} \\
\end{tabular}
\end{center}
\end{table*}

\subsection{Duration of the Stage A Superhumps}\label{sec:durationA}
The stage A superhumps had been observed for about three days in
 both MASTER J211258 and MASTER J203749. 

 The duration of the stage A (growing stage of 
the superhumps) is expected to be
 a good measure for the growth time of the 3:1 resonance.
 The growth rate of the 3:1 resonance is shown
 to be proportional to $q^2$ \citep{lub91SHa}, and
 a longer stage A is expected for a low $q$ system.
 This has recently been confirmed in a likely period bouncer
 ($q \lesssim 0.05$) SSS J122221.7$-$311523, which showed
a duration of stage A amounting to 11--15~d \citep{kat13j1222}.

 Although \citet{pat98egcnc} estimated the mass ratio of EG
 Cnc to be $q=0.027$, this value might be significantly too
 small for its short duration of stage A superhumps (stage A
 superhumps were not identified in this object; we estimated
 the duration of stage A to be less than 3~d from the epochs
 of early superhumps and fully grown superhumps).
 We propose that $q$ of EG Cnc should to be re-examined.

\subsection{Mass-Ratio from Stage A Superhumps}\label{sec:qfromstagea}

   Quite recently, \citet{osa13v344lyrv1504cyg}, proposed 
an important interpretation: the dynamical precession rate at
the 3:1 resonance is represented in the growing stage of superhumps
(stage A) when the eccentric wave is still confined to 
the location of the resonance. \citet{kat13qfromstageA}
 systematically studied stage A superhumps in SU UMa-type
dwarf novae and confirmed that the mass ratios determined
from fractional superhump excesses of stage A superhumps
are in good agreement with those determined from quiescent
eclipse observations or radial-velocity study.
\citet{kat13qfromstageA} also confirmed that the evolutionary
sequence using these mass ratios is in good agreement with
the most recent evolutionary sequence determined from
modern precise eclipse observations [such as \citet{lit08eclcv}].
Having the theoretical background \citep{osa13v344lyrv1504cyg},
we used this relation to estimate the mass ratio.

This relation gives mass ratios of MASTER J211258 and 
MASTER J203749 $q=0.081(2)$ and $q=0.097(8)$ respectively.

 The traditional method to estimate mass ratios from
 fractional superhump excesses of stage B superhumps, using
\begin{equation}
\epsilon=0.16(2)q+0.25(7)q^2
\label{eq:epsilon-q}
\end{equation}

\citep{Pdot} or

\begin{equation}
\epsilon=0.18q+0.29q^2
\label{eq:epsilon-q2}
\end{equation}

 \citep{pat05SH}, 
 gives mass ratios of MASTER J211258 and 
MASTER J203749 $q=0.054$ and $q=0.080$, respectively.

 The mass ratios from stage A superhumps are higher
 than the mass ratios from stage B superhumps.
 This discrepancy will be discussed in subsection \ref{sec:evolution}.

\subsection{Duration of the Early Superhump Stage}\label{sec:durationESH}

The early superhumps were observed for 12~d in 
MASTER J211258 and for 3~d in MASTER J203749 (we must note
that MASTER J203749 was not detected sufficiently early,
and this duration is a lower limit of the actual duration).
Table \ref{tab:earlysh} shows the duration of the early
 superhump stage of well-observed WZ Sge-type dwarf novae.
 
Early superhumps are thought to be a manifestation of the
 2:1 resonance \citep{osa02wzsgehump}. 
 During the stage of early superhumps,
 the 2:1 resonance is dominant
 since the development of
 the 2:1 resonance is expected to suppress the 3:1 resonance
 (\cite{lub91SHa}, \cite{osa03DNoutburst}).
 Therefore the duration of the early superhump stage
 is expected to be dependent on the disk mass at beginning
 of the outburst and the disk radius of 2:1 resonance
 relative to the Roche lobe.
 It is expected that small $q$ and large disk mass lead
 to long duration of the early superhump stage
 because the disk radius of 2:1 resonance is related to $q$.

In EG Cnc, this picture suggests that the comparatively
 short duration (less than 10~d, even if we consider
 the maximum duration of the observational gap) of
 the early superhump stage appears to be incompatible with
 the very small mass ratio $q=0.027$ estimated by {\citet{pat98egcnc}}.
 Either actual $q$ may be larger or the disk mass of EG Cnc 
 at the start of the superoutburst may have been 
 exceptionally small.

\begin{table*}
\caption{Duration of the early superhump stage.}
\label{tab:earlysh}
\begin{center}
\begin{tabular}{cccc}
\hline
Object & Year & Duration\commenta & References \\
\hline
EG Cnc & 1996 & 2--10 & \citet{mat96egcnciauc} \\
ASAS J0233 & 2006 & 8 & \citet{Pdot} \\
V455 And & 2007 & 7 & \citet{Pdot} \\
V466 And & 2008 & 11 & \citet{Pdot} \\
AL Com & 1995 & 5 & \citet{kat96alcom} \\
DV Dra & 2005 & ]5\commentb & \citet{Pdot} \\
BW Scl & 2011 & 10 & \citet{Pdot4} \\
WZ Sge & 2001 & 11 & \citet{ish02wzsgeletter} \\
OT J0120 & 2010 & 11 & \citet{Pdot2} \\
\hline
  \multicolumn{4}{l}{\commenta Unit d.} \\
  \multicolumn{4}{l}{\commentb "]" represents the lower limit.} \\
\end{tabular}
\end{center}
\end{table*}

\subsection{Period in Post-Superoutburst State}\label{sec:post}

\begin{figure}
  \begin{center}
    \FigureFile(88mm,110mm){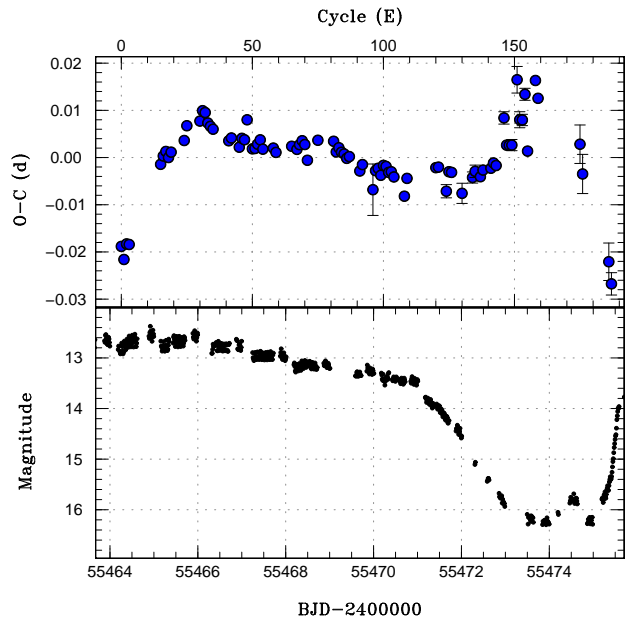}
  \end{center}
  \caption{The $O-C$ curve of EZ Lyn during the superoutburst
 [the data are taken from \citet{Pdot3}]. 
  (Upper:) O-C diagram of superhumps in EZ Lyn.
   An ephemeris of BJD 2455464.258+0.05966$E$ was used to draw this
   figure. (Lower:) Light curve. The data were binned to 0.01 d.}
  \label{fig:ocplot_j0804}
\end{figure}

During final part of the superoutburst, an increase in the superhump period
 was seen both in  MASTER J211258 and MASTER J203749.
The tendency is similar to that of EZ Lyn, the eclipsing
 WZ Sge-type dwarf nova with multiple rebrightenings \citep{Pdot3}.
The $O-C$ diagram of EZ Lyn is shown in figure \ref{fig:ocplot_j0804}.

\citet{lub92tilt} suggested that disk precession is due to a 
combination of effects of direct axisymmetric
 tidal forces from secondary, which is the most important,
 gas pressure in the eccentric mode and resonant wave stresses.
 Pressure forces act to decrease 
the precession rate. The tidal effect dominates for superhump binaries
 and produces a net prograde precession.
 The gas pressure effect produces a retrograde contribution
 and decrease the precession rate.
 The inward propagation of the eccentricity wave is accompanied by an increase
 of the pressure effect. Therefore the pressure effect is larger in
 stage B than in stage A. Furthermore, it is expected that
 the decrease in the pressure effect due to the transition to a cool
 state leads to the increase in the precession rate during
 final part of the superoutburst, which results in the increase in
 the superhump period.
 The increase in the period in MASTER J211258
 and MASTER J203749 during the final part of superoutburst
 is exactly what is expected for this interpretation.

 From \citet{kat13qfromstageA}, \citet{kat13j1222},
 the fractional superhump excesses (in frequency unit)
 of the stage A and post-superoutburst superhumps
 are expressed as follows:

\begin{equation}\label{eq:post}
\epsilon^*({\rm post})=Q(q)R(r_{\rm post})
\end{equation}

and

\begin{equation}\label{eq:stageA}
\epsilon^*({\rm stageA})=Q(q)R(r_{3:1}),
\end{equation}

where $r_{3 : 1}$ is the radius of 3:1 resonance

\begin{equation}\label{eq:3:1}
r_{3:1}=3^{(-2/3)}(1+q)^{-1/3}.
\end{equation}

 $\epsilon^*\equiv 1-P_{\rm orb}/P_{\rm SH}$, $r_{\rm post}$ is the
 dimensionless disk radius immediately after the outburst
 measured in units of binary separation $A$,

\begin{equation}\label{eq:post2}
Q(q)=\frac{1}{2} \frac{q}{\sqrt{1+q}},
\end{equation}

and
\begin{equation}
\label{eq:post3}
R(r) = \frac{1}{2}\frac{1}{\sqrt{r}} b_{3/2}^{(1)}(r),
\end{equation}
where
$\frac{1}{2}b_{s/2}^{(j)}$ is the Laplace coefficient
\begin{equation}
\label{eq:post4}
\frac{1}{2}b_{s/2}^{(j)}(r)=\frac{1}{2\pi}\int_0^{2\pi}\frac{\cos(j\phi)d\phi}
{(1+r^2-2r\cos\phi)^{s/2}}.
\end{equation}

Using the relation as follows:
\begin{equation}\label{eq:relation}
\frac{\epsilon^*{(\rm stageA)}}{\epsilon^*{(\rm post)}}
=\frac{R(r_{3:1})}{R(r_{\rm post})}
\end{equation}
 we can obtain $R(r_{\rm post})$.
 In MASTER J211258, post-superoutburst superhumps with a period
 of 0.06050(6) d were recorded in the final part
 of the superoutburst (BJD 2456124.5-2456126.5).
 The equation (\ref{eq:post}) gives $r=0.33$ using $q$ from stage A superhump.
 Kato et al. (2013) showed $0.30 \leq r \leq 0.38$ for
 the post-superoutburst state of WZ Sge-type dwarf novae.
 The agreement appears to be good.

\subsection{Post-Superoutburst Rebrightenings}\label{sec:rebrightenings}

\begin{figure}
  \begin{center}
    \FigureFile(88mm,110mm){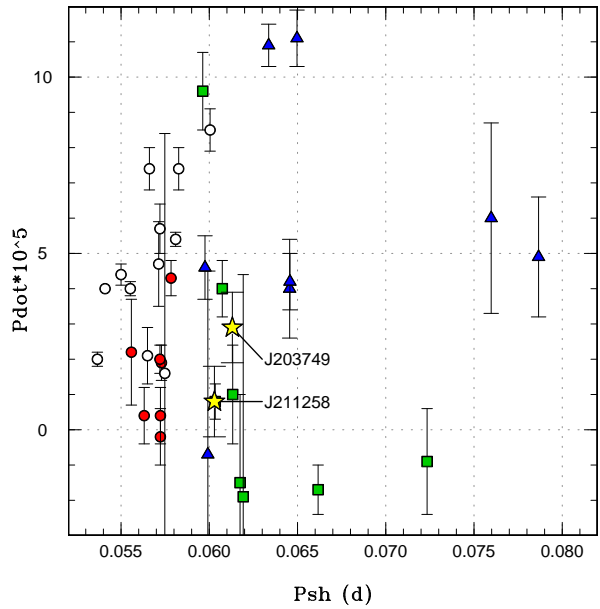}
  \end{center}
  \caption{$P_{\rm SH}$ versus $P_{\rm dot}$ for each type of rebrightenings.
 The star marks represent the two MASTER objects studied in this work.
 The other data are from \citet{Pdot}.
 The filled circles, filled squares,
 filled triangles, and open circles represent type-A, type-B, type-C and
 type-D, respectively. }
  \label{fig:pdot-psh}
\end{figure}

The rebrightenings of MASTER J211258 are considered
 to be type-B rebrightenings based on the classification by
 \citet{ima06tss0222}. Figure 
\ref{fig:pdot-psh}
 shows $P_{\rm dot}$ versus $P_{\rm SH}$, including the data reported in 
\citet{Pdot} and \citet{nak13j0120}. The filled circles, filled squares,
 filled triangles, and open circles represent type-A, type-B, type-C and
 type-D, respectively(these points represent different outbursts;
we hereafter call them ``type-X object'' assuminging that
the outburst type generally represents the property of the object).
 The locations of MASTER J211258 and MASTER J203749,
 expressed by filled stars, indicate that they
 are located close to the known type-B objects. 

Table \ref{tab:rebrightenings1} and \ref{tab:norebrightenings} list
 WZ Sge-type dwarf novae with multiple rebrightenings and those without
 a rebrightening, respectively.
Although two systems (QZ Ser and OGLE-GD-DN-014)
with longer superhump periods ($\geq 0.07$ d) and multiple
 rebrightenings have been reported, at least one of
them (QZ Ser) is an object which is not on the standard evolutionary 
track (Ohshima et al in prep.; \cite{tho02qzser}).
We therefore disregard the systems with longer
 superhump periods in table \ref{tab:rebrightenings1}.
   
\begin{figure}
  \begin{center}
    \FigureFile(80mm,70mm){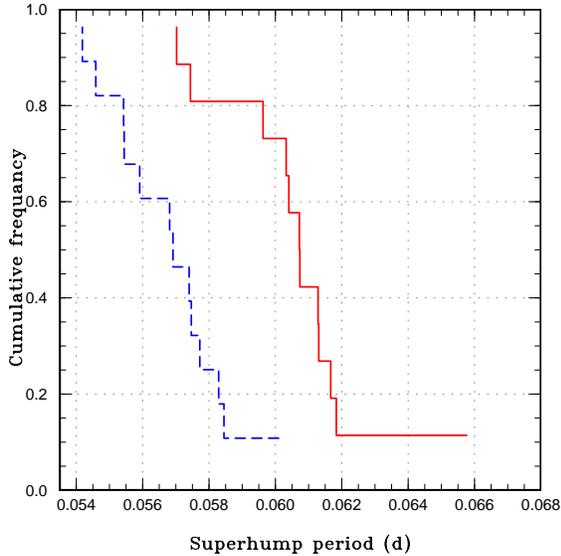}
  \end{center}
  \caption{Cumulative frequency of the distribution of
 superhump periods of 
WZ Sge-type stars with multiple rebrightenings (type-B) and
 those without a rebrightening (type-D).
Solid and dashed line indicate that of type-B objects
 and type-D objects, respectively.}
  \label{fig:reb_hist}
\end{figure}

 Figure \ref{fig:reb_hist} shows a cumulative frequency of the distribution
 of superhump periods. Solid and dashed line represent type-B
 and type-D objects, respectively.

The Kolmogorov-Smirnov test shows that the superhump periods 
distribution of type-D (no rebrightening) and type-B (multiple
 rebrightenings) do not belong to the same parent population
 with a significance level of $p<0.001$.
We can safely conclude that the superhump periods of type-B objects
 are longer than those of type-D objects.

\begin{table}
\caption{WZ Sge-type dwarf novae with multiple rebrightenings.
The data are from \citet{Pdot}, \citet{Pdot3}, \citet{Pdot4}
 and \citet{mro13OGLEDN2}.}\label{tab:rebrightenings1}
\begin{center}
\begin{tabular}{cccc}
\hline
Object & Year & $N_{reb}$\commenta & $P_{\rm SH}$ \\
\hline
UZ Boo & 1994 & 2: & 0.061743(38) \\
UZ Boo & 2003& 4 & 0.061922(33) \\
DY CMi & 2008 & 6 & 0.060736(9) \\
EG Cnc & 1996 & 6 & 0.060337(6) \\
AL Com & 2008 & ]4\commentb & 0.057174(6) \\
VX For & 2009 & 5 & 0.061327(12) \\
EZ Lyn & 2006 & 11 & 0.059537(31) \\
EZ Lyn & 2010 & 6 & 0.059630(16) \\
EL UMa & 2010 & ]4\commentb & 0.06045(6)\commentc \\
1RXS J0232 & 2007 & 4 & 0.066166(11) \\
OGLE-GD-DN-001 & 2007 & 4 & 0.06072(2) \\
MASTER J2037 & 2012 & ]3\commentb  & 0.061307(9) \\
MASTER J2112 & 2012 & 8  & 0.060291(4)  \\

\hline
  \multicolumn{4}{l}{\commenta Number of rebrightenings.} \\
 \multicolumn{4}{l}{\commentb "]" represents the lower limit.} \\
\multicolumn{4}{l}{\commentc  During the rebrightening phase.} \\
\end{tabular}
\end{center}
\end{table}

\begin{table}
\caption{WZ Sge-type dwarf novae without a rebrightening.}\label{tab:norebrightenings}
\begin{center}
\begin{tabular}{ccc}
\hline
Object & Year & $P_{\rm SH}$ \\
\hline
SV Ari & 2011 & 0.055524(14) \\
V455 And & 2007 & 0.057133(10) \\
V466 And & 2008 & 0.057203(15) \\
V592 Her & 1998 & 0.056498(13)  \\
V592 Her & 2010 & 0.056607(16) \\
V1108 Her & 2004 & 0.057480(34) \\
GW Lib & 2007 & 0.054095(10) \\
BW Scl & 2011 & 0.055000(8) \\
HV Vir & 2002 & 0.058266(17) \\
SDSS J1339 & 2011 & 0.058094(7) \\
OT J0238 & 2008 & 0.053658(7) \\
OT J2138 & 2010 & 0.055019(12) \\
OT J2109 & 2011 & 0.060045(26) \\

\hline
\end{tabular}
\end{center}
\end{table}

\begin{figure*}
 \begin{minipage}{0.5\hsize}
  \begin{center}
   \FigureFile(88mm,110mm){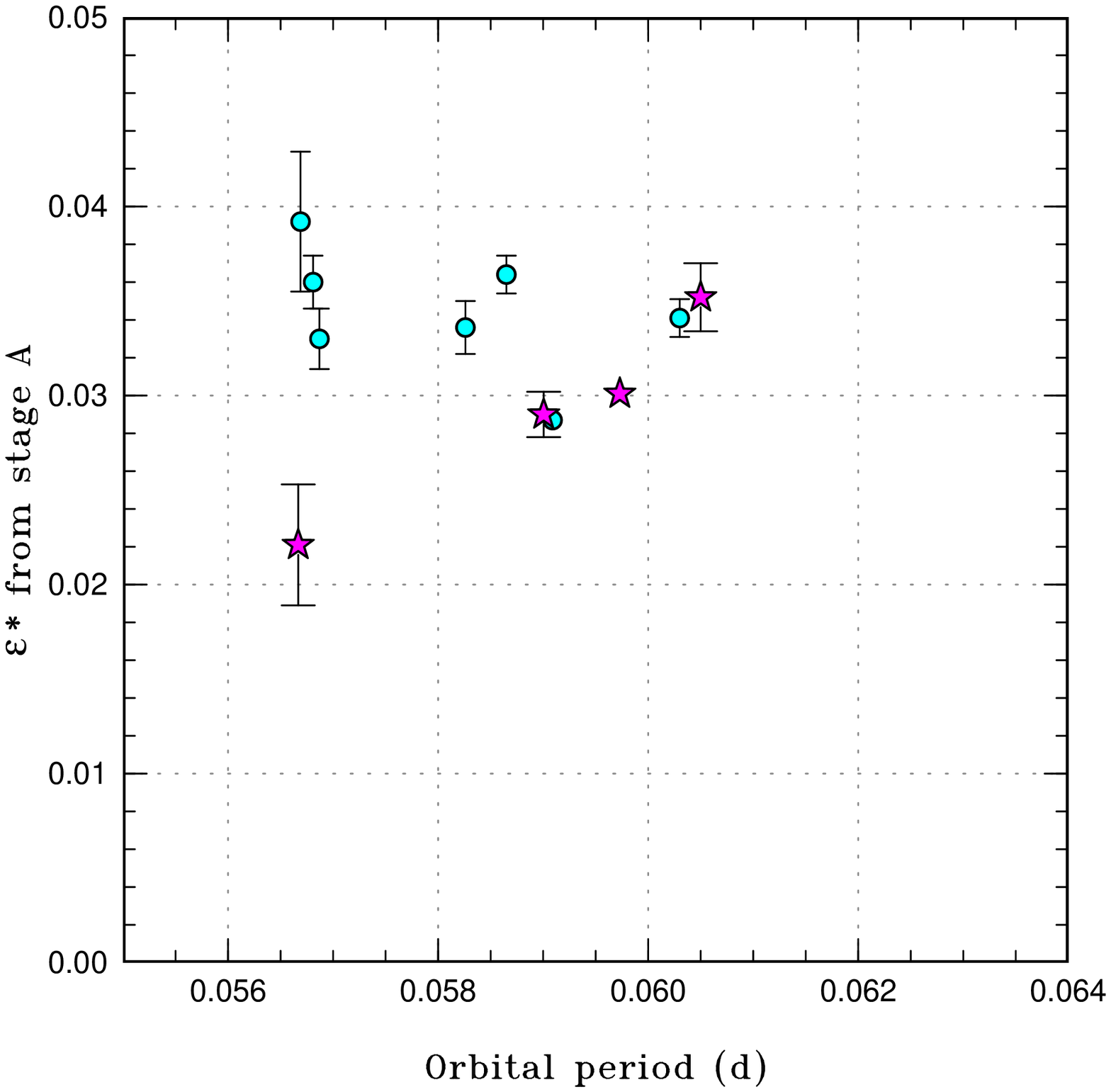}
  \end{center}
  \caption{$P_{\rm orb}$ versus $\epsilon^*$ for stage A superhumps.
 The data are from our result, \citet{Pdot}, \citet{Pdot3}, and \citet{Pdot4}.
 Filled circles and filled stars represent SU UMa-type
 dwarf novae without a rebrightening and WZ Sge-type dwarf novae
 with multiple rebrightenings, respectively.}
  \label{fig:epsilon_A}
 \end{minipage}
 \begin{minipage}{0.5\hsize}
  \begin{center}
   \FigureFile(88mm,110mm){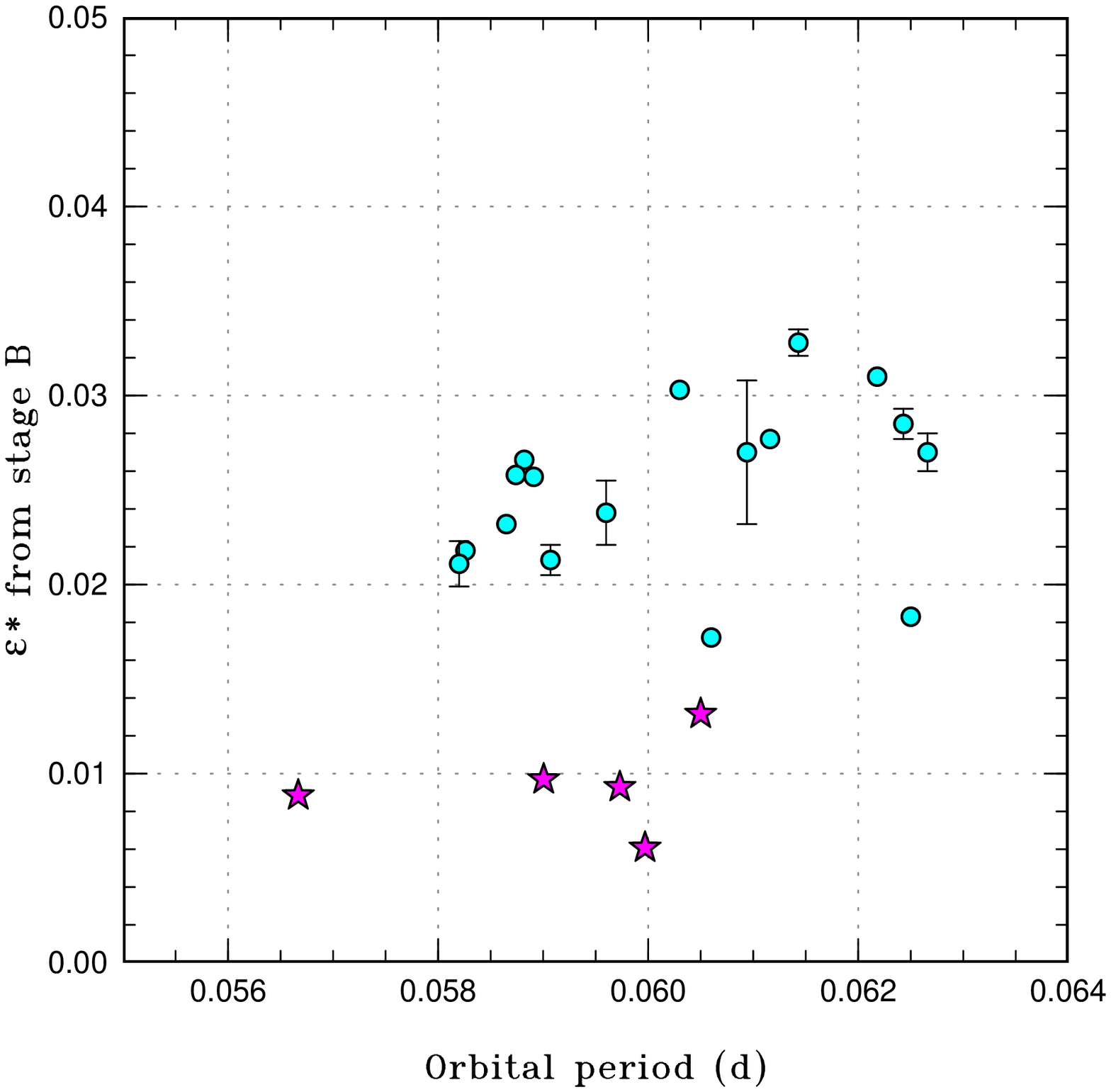}
  \end{center}
  \caption{$P_{\rm orb}$ versus $\epsilon^*$ for stage B superhumps.
 The data are from our result, \citet{Pdot}, \citet{Pdot3}, and \citet{Pdot4}.
 Filled circles and filled stars represent SU UMa-type dwarf novae
 without a rebrightening and WZ Sge-type dwarf novae
 with multiple rebrightenings, respectively.}
  \label{fig:epsilon_B}
 \end{minipage}
\end{figure*}

\begin{table*}
\caption{Fractional superhump excesses (in frequency unit) of
 WZ Sge-type dwarf novae with multiple rebrightenings.}
\label{tab:rebrightenings}
\begin{center}
\begin{tabular}{cccccc}
\hline
Object & $P_{\rm orb}$\commentc & $P_{\rm SH}$\commentc(stage B) & $\epsilon^*$ \commentb(stage B) & $P_{\rm SH}$\commentc(stage A) & $\epsilon^*$ \commentb(stage A)\\
\hline
UZ Boo & - & 0.061831(36) & - & 0.06354(24) & - \\
DY CMi & - & 0.060736(9) & - & - & - \\
EG Cnc & 0.05997 & 0.060337(6) & 0.00608(10) & - & - \\
AL Com & 0.056668 & 0.057174(6) & 0.00885(11) & 0.05795(18) & 0.0221(32)\\
VX For & - & 0.061327(12) & - & - & - \\
EZ Lyn & 0.059005 & 0.059584(24) & 0.00970(41) & 0.06077(7) & 0.0290(12) \\
1RXS J0232 & - & 0.066166(11) & - & - & - \\
MASTER J2037 & 0.06051 & 0.061307(9) & 0.01316(15) & 0.06271(11) & 0.0352(18) \\
MASTER J2112 & 0.05973 & 0.06029(4) & 0.00929(67) & 0.06158(5) & 0.0301(8)\\

\hline
\multicolumn{4}{l}{\commentc Unit d.} \\
\multicolumn{4}{l}{\commentb Fractional superhump excess (in frequency unit).} \\
\end{tabular}
\end{center}
\end{table*}

\begin{table}
\caption{Fractional superhump excesses (in frequency unit)
 from stage A superhumps
 for SU UMa-type dwarf novae without a rebrightening 
with the period range of  0.05875--0.06458 d.}
 \label{tab:suuma_stageA}
\begin{center}
\begin{tabular}{cccc}
\hline
Object & $P_{\rm orb}$\commentc & $P_{\rm SH}$\commentc (stage A) & $\epsilon^*$ \commentb \\
\hline
V1040 Cen & 0.06030 & 0.06243(6) & 0.0341(10) \\
WX Cet & 0.05826 & 0.06029(8) & 0.0336(14) \\
PU CMa & 0.05669 & 0.05901(21) & 0.0392(37) \\
MM Hya & 0.05759 & 0.06163(24) &  0.0656(42) \\
SW UMa & 0.05681 & 0.05893(8) &  0.0360(14) \\
SDSS J1610 & 0.05687 & 0.05881(9) &  0.0330(16) \\
OT J1044 & 0.05909 & 0.06084(2) &  0.0287(3) \\
OT J2109 & 0.05865 & 0.06087(6) & 0.0364(10) \\
\hline
\multicolumn{4}{l}{\commentc Unit d.} \\
\multicolumn{4}{l}{\commentb Fractional superhump excess (in frequency unit).} \\
\end{tabular}
\end{center}
\end{table}

\begin{table}
\caption{Fractional superhump excesses (in frequency unit)
 from stage B superhumps
 for SU UMa-type dwarf novae without a rebrightening 
 with the period range of 0.05875--0.06458 d.}\label{tab:suuma_stageB}
\begin{center}
\begin{tabular}{cccc}
\hline
Object & $P_{\rm orb}$\commentc & $P_{\rm SH}$\commentc (stage B) & $\epsilon^*$ \commentb \\
\hline
NZ Boo & 0.05891 & 0.06046(1) & 0.0257(2) \\
V436 Cen & 0.06250 & 0.06366(1) & 0.0183(2) \\
V1040 Cen & 0.06030 & 0.06218(3) & 0.0303(5) \\
WX Cet & 0.05826 & 0.05956(3) &  0.0218(5) \\
HO Del & 0.06266 & 0.06440(6) & 0.0270(10) \\
XZ Eri & 0.06116 & 0.06285(3) & 0.0277(5) \\
AQ Eri & 0.06094 & 0.06256(23) & 0.0270(38) \\
V2051 Oph & 0.06243 & 0.06426(5) & 0.0285(8) \\
V1159 Ori & 0.06218 & 0.06417(3) & 0.0310(5) \\
V4140 Sgr & 0.06143 & 0.06351(4) & 0.0328(7) \\
QZ Vir & 0.05882 & 0.06043(2) & 0.0266(3) \\
SDSS J0903 & 0.05907 & 0.06036(5) & 0.0213(8) \\
SDSS J1250 & 0.05874 & 0.06030(2) & 0.0258(3) \\
SDSS J2048 & 0.06060 & 0.06166(2) & 0.0172(3) \\
OT J105122 & 0.05960 & 0.06105(10) & 0.0238(17) \\
OT J1706 & 0.05820 & 0.05946(7) & 0.0211(12) \\
OT J2109 & 0.05865 & 0.06005(2) & 0.0232(3) \\
\hline
\multicolumn{4}{l}{\commentc Unit d.} \\
\multicolumn{4}{l}{\commentb Fractional superhump excess (in frequency unit).} \\
\end{tabular}
\end{center}
\end{table}

\subsection{Evolutionary State}\label{sec:evolution}

Table \ref{tab:rebrightenings} lists fractional superhump excesses
 (in frequency unit) of
 WZ Sge-type dwarf novae with multiple rebrightenings (type-B).
 We find the $95\%$ quantile of superhump periods of
 WZ Sge-type dwarf novae with multiple rebrightenings between 0.05875 and
 0.06458~d (figure \ref{fig:reb_hist}).
 In this period region, there are many SU UMa-type stars without
 a rebrightening.  We therefore make a comparison between them and
 WZ Sge-type dwarf novae with multiple rebrightenings.
 The fractional superhump excesses (in frequency unit)
 from stage A superhumps and stage B
 superhumps are also listed in table \ref{tab:suuma_stageA} and
 table \ref{tab:suuma_stageB}, respectively.
 
Figure \ref{fig:epsilon_A} and figure \ref{fig:epsilon_B} show
 the relation between $P_{\rm orb}$ versus $\epsilon^*$
 for WZ Sge-type dwarf novae with multiple rebrightenings
 (listed in table \ref{tab:rebrightenings})
 and SU UMa-type dwarf novae (listed in tables \ref{tab:suuma_stageA}
 and \ref{tab:suuma_stageB}).

 Figure \ref{fig:epsilon_A} shows $\epsilon^*$ from
 stage A superhumps of SU UMa-type
 dwarf novae without a rebrightening (filled circles) and
 WZ Sge-type dwarf novae with multiple rebrightenings
 (filled stars).
 Figure \ref{fig:epsilon_B} shows $\epsilon^*$ from
 stage B superhumps of SU UMa-type
 dwarf novae without a rebrightening (filled squares) and
 WZ Sge-type dwarf novae with multiple rebrightenings
 (filled triangles).

 $\epsilon^*$ from stage A superhumps of WZ Sge-type dwarf novae
 with multiple rebrightenings are located in a region with a little
 smaller $\epsilon^*$ than those of SU UMa-type dwarf novae
 without a rebrightening.
 Using the relation between $\epsilon^*$ of stage A superhumps
 and $q$ (subsection \ref{sec:qfromstagea}), $q$ of WZ Sge-type dwarf novae
 with multiple rebrightenings are estimated close to (only slightly
 smaller than)
 those of SU UMa-type dwarf novae without a rebrightening.
 However, they are two different populations using $\epsilon^*$ from stage
 B superhumps. $\epsilon^*$ from stage B superhumps of WZ Sge-type
 dwarf novae with multiple rebrightenings are clearly smaller
 than that of SU UMa-type dwarf novae without a rebrightening.

 \citet{mur00SHprecession} formulated the hydrodynamical
 precession $\omega$ in the form:

\begin{equation}
\omega=\omega_{\rm dyn} + \omega_{\rm pres}
\end{equation} 

where $\omega_{\rm dyn}$ and $\omega_{\rm pres}$ are the dynamic
 precession and the pressure contribution to the precession.
 $\omega_{\rm pres}/\omega_{\rm orb}$ corresponds to 
 $\epsilon_{\rm A}^*-\epsilon_{\rm B}^*$,
 where $\epsilon_{\rm A}^*$ and $\epsilon_{\rm B}^*$ are
 $\epsilon^*$ from stage A
 and that of stage B, respectively.
 $\epsilon_{\rm A}^*-\epsilon_{\rm B}^*$ is generally
 similar between those two types of systems.
 It is shown that $\epsilon_{\rm A}^*-\epsilon_{\rm B}^*\simeq 0.010$ 
 in SU UMa-type dwarf novae without a rebrightening and $\simeq 0.015$
 in WZ Sge-type dwarf novae with multiple rebrightenings.

 We consider two possible interpretations of this result
 of great interest.
 The first interpretation, $q$ from stage B superhumps is
 indeed the case and that $q$ from stage A does not reflect
 the true $q$.
 The second interpretation,
 $q$ from stage B superhumps does not reflect the true $q$.
 If the former interpretation is correct, the small
 $q$ of WZ Sge-type dwarf novae with multiple rebrightenings
 suggest that the systems are good candidates for period bouncers.
 If the latter interpretation is the case, there is insignificant difference 
 in $q$ between WZ Sge-type dwarf novae with multiple
 rebrightenings and SU UMa-type dwarf novae without a rebrightening.
 In the latter interpretation, small $q$ from stage B superhumps
 may be explained as a result of a stronger pressure effect
 during stage B.
 These two possibilities should be tested by future studies.
 We, however, have no idea why growing superhumps arise from
 a radius other than the 3:1 resonance, and  
 we are in favor of the latter interpretation. 

\begin{figure}
  \begin{center}
    \FigureFile(90mm,105mm){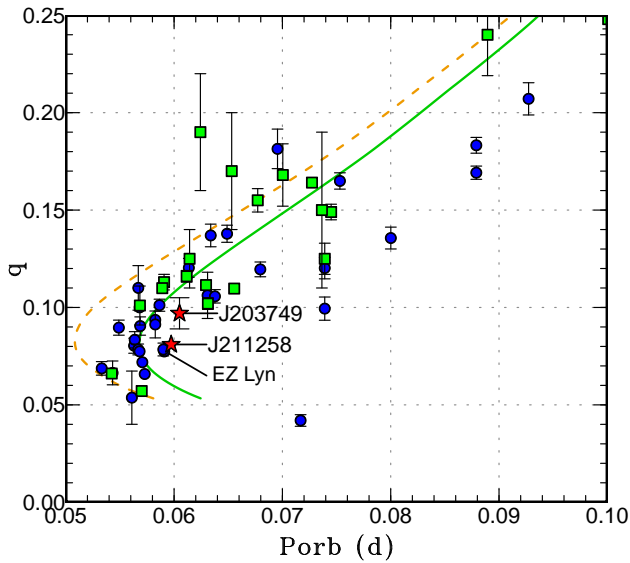}
  \end{center}
  \caption{$q$ versus $P_{\rm orb}$. The data are from \citet{kat13qfromstageA}.
The filled circles, filled squares and filled stars represent
$q$ from stage A superhumps, $q$ measured by eclipse and
 our data ($q$ from stage A superhumps), respectively. Our data are
 located in almost the same place as other SU UMa-type dwarf novae.
 The dash curved line and solid curved line represent evolutionary
 track of the standard evolutional theory and
 that of the modified evolutional theory \citep{kni11CVdonor}, respectively.}
  \label{fig:evol}
\end{figure}

 Our result suggests a new picture that both classes of objects
 are located in the same place in the evolutionary trace
 (figure \ref{fig:evol}).
 The difference between these classes may be a result of
 different mass transfer rates.
 This possibility should be explored by future studies.

\section{Summary}\label{sec:summary}
We obtained  photometric observations of WZ Sge-type dwarf novae,
 MASTER OT J211258.65$+$242145.4 and MASTER OT J203749.39+552210.3
 in 2012.
 The early superhumps with a period of 0.059732(3) d (MASTER J211258)
 and of 0.06050 d (MASTER J203749) were recorded.
 During superoutburst, ordinary superhumps with a period of 0.060291(4) d
 (MASTER J211258) and of 0.061307(9) d (MASTER J203749) were
 observed.
 The $O-C$ curve clearly showed the presence of stage A and stage B.
  Both dwarf novae exhibited multiple rebrightenings after the main
 superoutburst.

 In both dwarf novae, the increase of brighteness were seen during
 the evolutionary stage of ordinary superhumps.
 Compared with other systems showing this phenomenon, we can see a
 possible tendency that those systems exhibit
 long rebrightening or multiple rebrightenings.

 From the distribution of superhump periods, 
 those superhump periods of type-B stars
 are systematically longer than those of type-D stars.

 We estimated binary mass ratios by a new method using the period
 of superhumps in SU UMa-type dwarf novae during stage A superhumps.
 The method gave higher mass ratios for these objects than those
 from stage B superhumps by the traditional method.
This suggests that the stage B superhump period is
 shorter due to effects of gas pressure.
 This result leads to a different picture of the evolutionary state in
 WZ Sge-type dwarf novae with multiple rebrightenings although
 they have been expected to be good candidates for
 period bouncers due to the small mass ratios expected
 from the period of stage B superhumps.

 In the final part of the superoutburst, the increase in the superhump
 periods was seen in both systems.
 We presented an interpretation that the increase in the 
 superhump periods is a result of the decrease
 of pressure effect during the final stage of the superoutburst and showed
 that the derived disk radius is in fair agreement
 with the previous work.

\medskip

We acknowledge with thanks the variable star
observations from the AAVSO International Database contributed by
observers worldwide and used in this research.
This work is deeply indebted to outburst detections and quick
announcement by the MASTER network.
We are grateful to many amateur observers.


\end{document}